%% file: UIREucPoincv7.tex
\documentclass[12pt]{article}
\usepackage{amsmath, amsthm}
\usepackage{amscd}
\usepackage{amssymb}
\usepackage{array}
\usepackage{epsfig}
\usepackage{color}
\usepackage{graphicx}
\usepackage{enumitem}
\usepackage[hidelinks]{hyperref}

\newtheorem{thm}{Theorem}[section]
\newtheorem{theorem}[thm]{Theorem}

\theoremstyle{definition}
\newtheorem{definition}[thm]{Definition}
\newtheorem{example}[thm]{Example}

\begin{document}

\newcommand{\comment}[1]{{\color{blue}\rule[-0.5ex]{2pt}{2.5ex}}
\marginpar{\scriptsize\begin{flushleft}\color{blue}#1\end{flushleft}}}

\newcommand{\rojo}{\color{red}}
\newcommand{\azul}{\color{blue}}

\newcommand{\be}{\begin{equation}}
\newcommand{\ee}{\end{equation}}
\newcommand{\beq}{\begin{equation}}
\newcommand{\eeq}{\end{equation}}
\newcommand{\ba}{\begin{align}}
\newcommand{\ea}{\end{align}}
\newcommand{\ban}{\begin{align*}}
\newcommand{\ean}{\end{align*}}
\newcommand{\ds}{\displaystyle}

\newcommand{\cid}{\relax{\rm 1\kern-.28em 1}}
\newcommand{\id}{\mathrm{id}}
\newcommand{\R}{\mathbb{R}}
\newcommand{\N}{\mathbb{N}}
\newcommand{\C}{\mathbb{C}}
\newcommand{\Z}{\mathbb{Z}}
\newcommand{\bH}{\mathbb{H}}
\newcommand{\g}{\mathfrak{G}}
\newcommand{\e}{\epsilon}

\newcommand{\hs}{\hfill\square}
\newcommand{\hbs}{\hfill\blacksquare}

\newcommand{\bp}{\mathbf{p}}
\newcommand{\bmax}{\mathbf{m}}
\newcommand{\bT}{\mathbf{T}}
\newcommand{\bU}{\mathbf{U}}
\newcommand{\bP}{\mathbf{P}}
\newcommand{\bA}{\mathbf{A}}
\newcommand{\bm}{\mathbf{m}}
\newcommand{\bIP}{\mathbf{I_P}}

\newcommand{\fso}{\mathfrak{so}}

\newcommand{\cL}{\mathcal{L}}
\newcommand{\cA}{\mathcal{A}}
\newcommand{\cB}{\mathcal{B}}
\newcommand{\cC}{\mathcal{C}}
\newcommand{\cI}{\mathcal{I}}
\newcommand{\cO}{\mathcal{O}}
\newcommand{\cG}{\mathcal{G}}
\newcommand{\cJ}{\mathcal{J}}
\newcommand{\cF}{\mathcal{F}}
\newcommand{\cP}{\mathcal{P}}
\newcommand{\ep}{\mathcal{E}}
\newcommand{\E}{\mathcal{E}}
\newcommand{\cH}{\mathcal{H}}
\newcommand{\cPO}{\mathcal{PO}}
\newcommand{\cl}{\ell}
\newcommand{\cFG}{\mathcal{F}_{\mathrm{G}}}
\newcommand{\cHG}{\mathcal{H}_{\mathrm{G}}}
\newcommand{\Gal}{G_{\mathrm{al}}}
\newcommand{\cQ}{G_{\mathcal{Q}}}
\newcommand{\cT}{\mathcal{T}}
\newcommand{\cM}{\mathcal{M}}
\newcommand{\cR}{\mathcal{R}}
\newcommand{\cS}{\mathcal{S}}
\newcommand{\qq}{\mathbf{q}}

\newcommand{\ri}{\mathrm{i}}
\newcommand{\re}{\mathrm{e}}
\newcommand{\rd}{\mathrm{d}}
\newcommand{\rSt}{\mathrm{St}}
\newcommand{\rGL}{\mathrm{GL}}
\newcommand{\rSU}{\mathrm{SU}}
\newcommand{\rSL}{\mathrm{SL}}
\newcommand{\rSO}{\mathrm{SO}}
\newcommand{\rOSp}{\mathrm{OSp}}
\newcommand{\rSpin}{\mathrm{Spin}}
\newcommand{\rsl}{\mathrm{sl}}
\newcommand{\rM}{\mathrm{M}}
\newcommand{\rU}{\mathrm{U}}
\newcommand{\rdiag}{\mathrm{diag}}
\newcommand{\rP}{\mathrm{P}}
\newcommand{\rdeg}{\mathrm{deg}}
\newcommand{\rStab}{\mathrm{Stab}}
\newcommand{\rcof}{\mathrm{cof}}
\newcommand{\rAut}{\mathrm{Aut}}
\newcommand{\rInd}{\mathrm{Ind}}
\newcommand{\rISO}{\mathrm{ISO}}
\newcommand{\rEnd}{\mathrm{End}}

\newcommand{\M}{\mathrm{M}}
\newcommand{\End}{\mathrm{End}}
\newcommand{\Hom}{\mathrm{Hom}}
\newcommand{\diag}{\mathrm{diag}}
\newcommand{\rspan}{\mathrm{span}}
\newcommand{\rank}{\mathrm{rank}}
\newcommand{\Gr}{\mathrm{Gr}}
\newcommand{\ber}{\mathrm{Ber}}

\newcommand{\fo}{\mathfrak{o}}
\newcommand{\fsl}{\mathfrak{sl}}
\newcommand{\fg}{\mathfrak{g}}
\newcommand{\ff}{\mathfrak{f}}
\newcommand{\fgl}{\mathfrak{gl}}
\newcommand{\fosp}{\mathfrak{osp}}
\newcommand{\fm}{\mathfrak{m}}

\newcommand{\ttau}{\tilde\tau}

\newcommand{\str}{\mathrm{str}}
\newcommand{\Sym}{\mathrm{Sym}}
\newcommand{\tr}{\mathrm{tr}}
\newcommand{\defi}{\mathrm{def}}
\newcommand{\Ber}{\mathrm{Ber}}
\newcommand{\spec}{\mathrm{Spec}}
\newcommand{\sschemes}{\mathrm{(sschemes)}}
\newcommand{\sschemeaff}{\mathrm{ {( {sschemes}_{\mathrm{aff}} )} }}
\newcommand{\rings}{\mathrm{(rings)}}
\newcommand{\Top}{\mathrm{Top}}
\newcommand{\sarf}{ \mathrm{ {( {salg}_{rf} )} }}
\newcommand{\arf}{\mathrm{ {( {alg}_{rf} )} }}
\newcommand{\odd}{\mathrm{odd}}
\newcommand{\alg}{\mathrm{(alg)}}
\newcommand{\sa}{\mathrm{(salg)}}
\newcommand{\sets}{\mathrm{(sets)}}
\newcommand{\SA}{\mathrm{(salg)}}
\newcommand{\salg}{\mathrm{(salg)}}
\newcommand{\varaff}{ \mathrm{ {( {var}_{\mathrm{aff}} )} } }
\newcommand{\svaraff}{\mathrm{ {( {svar}_{\mathrm{aff}} )}  }}
\newcommand{\ad}{\mathrm{ad}}
\newcommand{\Ad}{\mathrm{Ad}}
\newcommand{\pol}{\mathrm{Pol}}
\newcommand{\Lie}{\mathrm{Lie}}
\newcommand{\Proj}{\mathrm{Proj}}
\newcommand{\rGr}{\mathrm{Gr}}
\newcommand{\rFl}{\mathrm{Fl}}
\newcommand{\rPol}{\mathrm{Pol}}
\newcommand{\rdef}{\mathrm{def}}
\newcommand{\rE}{\mathrm{E}}
\newcommand{\rId}{\mathrm{Id}}
\newcommand{\rO}{\mathrm{O}}

\newcommand{\cCl}{\mathcal{C}l}

\newcommand{\sym}{\cong}
\newcommand{\al}{\alpha}
\newcommand{\lam}{\lambda}
\newcommand{\de}{\delta}
\newcommand{\D}{\Delta}
\newcommand{\s}{\sigma}
\newcommand{\lra}{\longrightarrow}
\newcommand{\ga}{\gamma}
\newcommand{\ra}{\rightarrow}

\newcommand{\tit}{t}
\newcommand{\ts}{\tilde{s}}
\newcommand{\ty}{\tilde{y}}
\newcommand{\titp}{{t}^{\, '}}
\newcommand{\hv}{ \hat{t}}
\newcommand{\hu}{\hat{\tau}}
\newcommand{\hy}{\hat{y}}
\newcommand{\hS}{\hat{S}(x)}
\newcommand{\hx}{\hat{x}}
\newcommand{\hT}{\hat{T}}

\newcommand{\vb}{\mathbf{b}}
\newcommand{\vp}{\mathbf{p}}
\newcommand{\vq}{\mathbf{q}}
\newcommand{\vcq}{\mathbf{Q}}
\newcommand{\vx}{\mathbf{x}}

\newcommand{\var}{\varepsilon}


\hyphenation{coun-ta-ble}
\hyphenation{pro-duct}
\hyphenation{ge-ne-ra-tors}
\hyphenation{ne-ga-tive}

\smallskip
    \centerline{\LARGE \bf  On the Unitary Irreducible }\medskip\centerline{\LARGE \bf Representations   of the $D=2$ }\medskip\centerline{\LARGE \bf Euclidean and  Poincar\'{e} Groups}
\vskip 1cm
\centerline{ G. Camilletti$^{1}$, M. A. Lled\'{o}$^{1}$, M. A del Olmo$^{2}$} \vskip 0.5cm

\centerline{\it $^1$
Departament de F\'{\i}sica Te\`{o}rica, Universitat de Val\`{e}ncia (UVEG) and}

\centerline{\it IFIC(CSIC-UVEG) - Instituto de F\'isica Corpuscular.}\centerline{\it  C/ Vicent Andr\'es Estell\'es, 19. 46100 Burjassot.  Spain.}

\medskip

\centerline{\it $^2$
Departamento de F\'isica Te\'orica, At\'omica y \'Optica and }

\centerline{\it IMUVA (Mathematical Research Institute)}
\centerline{\it Universidad de Valladolid,} \centerline{\it Paseo Bel\'en, 7. 47011 Valladolid, Spain.}

\bigskip

\centerline{{\footnotesize e-mail:
giovanni.camilletti@uv.es, maria.lledo@ific.uv.es, marianoantonio.olmo@uva.es}}

\vskip 0.5cm

\begin{abstract}

We present an explicit construction of the unitary irreducible
representations of the two-dimensional Euclidean and Poincar\'e groups,
together with their Spin double covers, by means of Mackey's theory of
induced representations for semidirect products. In dimension $D=2$, the
simplicity of the corresponding little groups allows a complete explicit
treatment of momentum orbits, equivariant wavefunctions, and representation
operators.

For the Euclidean group, the matrix elements of the in\-fi\-nite-di\-men\-sion\-al
representations are expressed in terms of Bessel functions. For the
Poincar\'e group, the richer Lorentzian orbit structure leads to matrix
elements involving modified Bessel and Hankel functions and, in some cases,
tempered distributions, requiring the use of Rigged Hilbert Spaces.

This work illustrates the interplay among induced representations, harmonic
analysis on Lie groups, Spin geometry, and special functions in a fully
explicit relativistic setting.

\end{abstract}


\hyphenation{ana-ly-sis}

\newpage

\section{Introduction}
\label{sec:Introduction}

The theory of unitary irreducible representations of Lie groups occupies a
central position in modern mathematical physics. In particular, the
representation theory of the Poincar\'{e} group provides the group-theoretical
framework underlying the description of elementary particles and relativistic
quantum systems. 
The unitary irreducible representations of the Poincar\'{e} group play a central role in relativistic quantum theory. In his seminal work  \cite{wi}, Wigner showed that elementary particles can be classified according to the unitary irreducible representations of the Poincar\'{e} group, obtained by inducing representations from the stabilizer (or little group) of momentum orbits. This construction was later placed in a systematic mathematical framework by Mackey through his theory of induced representations for locally compact groups \cite{ma1,ma2,ma3,BarutRaczka1986}, which provides a general method for semidirect product groups. Within this approach, the representation theory of the Poincar\'{e} group arises, naturally, from the action of the Lorentz group on momentum space and from the associated little groups.

Mackey's theory constitutes one of the fundamental tools in harmonic
analysis and representation theory, since it provides a systematic procedure
for constructing unitary irreducible representations from the action of a
group on the dual of a normal abelian subgroup. In the context of
mathematical physics, this approach is particularly relevant for spacetime
symmetry groups, where physically meaningful quantities such as mass and
spin arise naturally from the orbit structure and the representations of the
corresponding little groups.  These geometric
aspects also play an important role in gauge approaches to gravity and
relativistic symmetry structures \cite{Tresguerres2013}. The present work also fits naturally within the
broader geometric approach to representation theory, closely related to
orbit methods, harmonic analysis on Lie groups, and geometric quantization
\cite{Kirillov2004,AliEnglis2005}.

 Although the classification of the unitary irreducible representations of
the Poincar\'e group is well known in arbitrary dimension (for the  case of 1+3  dimensions see \cite{wi,bw}), explicit
realizations are rarely presented in full detail. In higher dimensions, the
structure of the little groups and the associated Wigner rotations rapidly
become technically involved, making explicit computations difficult. As a
consequence, most treatments focus mainly on classification results,
Casimir operators, or abstract induced constructions. Hence, explicit realizations —particularly at the level of matrix elements of the representation operators— are rarely written down in full detail.

 The purpose of the present work is to provide a complete and explicit
implementation of Mackey's construction for the Euclidean and Poincar\'e
groups in spacetime dimension $D=2$. Despite their apparent simplicity,
these groups provide particularly instructive examples because all the
ingredients of the induced representation program can be carried out
explicitly: the orbit decomposition in momentum space, the determination of
invariant measures, the construction of equivariant functions, the
realization of infinitesimal generators, and the explicit computation of the
matrix elements of the representation operators.

 The two-dimensional case is also of independent physical and mathematical
interest. Quantum field theories in $1+1$ dimensions constitute an important
laboratory for exactly solvable and integrable models
\cite{co,zz}, where the representation theory of
the Poincar\'e group governs the structure of one-particle states. On the
other hand, the Euclidean group $E(2)$ appears naturally as the little group
associated with massless particles in four-dimensional relativistic theories.
Its unitary irreducible representations include, not only the familiar
helicity representations, but also the continuous-spin representations
originally identified by Wigner and later studied in several contexts
\cite{bm,Brink2002,bs,Schuster2013,glvv,st}.

\hyphenation{pass-ing}
An additional aspect of the present work is the systematic treatment of the
double covers of the orthogonal groups involved. Since relativistic quantum
theory naturally requires both bosonic and fermionic representations, it is
convenient to formulate the construction in terms of the corresponding Spin
groups and Clifford algebras. In the two-dimensional setting, the little
groups reduce essentially to either the identity or $\mathbb{Z}_2$, allowing
the complete representation-theoretical construction to be performed
explicitly.

For the Euclidean group and its double cover, we determine the irreducible
representations associated with both trivial and non-trivial momentum
orbits, and compute explicitly the matrix elements of the infinite-dimensional
representations in a basis adapted to the rotation generator. These matrix
elements are expressed in terms of Bessel functions, revealing the close
relation between harmonic analysis on Lie groups and classical special
functions \cite{Vilenkin1991}.

For the Poincar\'e group in $1+1$ dimensions, the orbit structure is richer
due to the Lorentzian signature of momentum space. We classify all the
relevant orbits and construct the associated induced representations
explicitly. In particular, we obtain concrete expressions for the matrix
elements of the representation operators in terms of Bessel, modified
Bessel, and Hankel functions. Some of these matrix elements naturally
acquire a distributional character, requiring the use of Rigged Hilbert
Spaces and generalized eigenvectors
\cite{gelfand,bohm-gadella,olmo2016,olmo2019}.

 From a broader perspective, the present work may be regarded as a bridge
between the abstract formulation of induced representation theory and
explicit computational realizations relevant to mathematical physics. It
also illustrates the interplay among Lie groups, harmonic analysis, Clifford
algebras, special functions, and geometric methods in representation theory,
in a setting where all computations can be performed explicitly.
\hyphenation{pers-pec-tives}
\hyphenation{rea-li-za-tions}

\hyphenation{pro-ducts} 

The paper is organized as follows. In Section~ \ref{sec:Mackey}  we review the basic
in\-gre\-di\-ents of induced representation theory and state Mackey's theorem for
semidirect products, fixing the notation used throughout the paper.
Section~\ref{euclidean-sec}  introduces the two-dimensional Euclidean group and its double
cover. In Section~\ref{sec:UIR(2)}  we construct explicitly the unitary irreducible
representations of $\rE(2)$ and compute their matrix elements. Section~\ref{sec:Poincare}
reviews the structure of the two-dimensional Poincar\'e group and emphasizes
its differences with respect to the Euclidean case. Section~\ref{sec:uir(1,1)} contains the
explicit construction of its unitary irreducible representations and the
analysis of the associated matrix elements. Finally, the appendices collect
the necessary material on Clifford algebras, Spin groups, and Rigged Hilbert
Spaces.

\hyphenation{sys-te-ma-tic}

\section{Mackey's theorem for unitary representations of semidirect products}\label{sec:Mackey}\hyphenation{re-pre-sen-ta-tions}

 The Mackey theory of induced representations \cite{ma1,ma2,ma3}  is fundamental in both mathematics and physics, because it provides a rigorous and systematic framework for constructing and classifying unitary irreducible representations of non-trivial groups, starting from those of certain subgroups. In mathematics, it plays a central role in both the representation theory of locally compact groups and in harmonic analysis, where it clarifies the structure of group actions, orbit spaces, and invariant measures. In physics, Mackey's method is essential for the group-theoretical classification of quantum states, particularly in relativistic quantum mechanics and quantum field theory, where representations of the Poincar\'e group, obtained via induction, encode physically 
 observable properties such as mass and spin.
 
 In this section we introduce the pieces that we need from the theory of induced representations and, at the end, we will state the theorem of Mackey for semidirect products. This also serves to establish the notation used throughout the paper.
\hyphenation{sys-te-ma-tic}
\subsection{Induced representations}\label{inducederepresentations}

   Let $H\subset G$ be a closed subgroup of the group $G$, a locally compact, second countable group, equipped with a left-invariant measure. We consider a unitary representation of $H$, $\sigma: H\rightarrow \rAut(\cH_\sigma)$, where $\cH_\sigma$ is the Hilbert space supporting the representation $\sigma$, with inner product $\langle\,\cdot\,,\,\cdot\,\rangle_\sigma$. We also denote as $G/H$ the coset space,  and as $p: G\rightarrow G/H$, the natural projection.

   On  the set $G\times \cH_\sigma$ we define the equivalence relation
   $$
   (g,v)\sim(g',v')\quad \hbox{if}\quad g'=gh \quad \hbox{and}\quad v'=\sigma(h^{-1})v\quad \hbox{for some } h\in H.
   $$
   The quotient set is denoted as
   \be\label{bundle}
   E=G\times_H\cH_\sigma,
 \ee  
    and it is a fiber bundle over $G/H$, with fiber $\cH_\sigma$. On $G\times \cH_\sigma$ there is an obvious action of $G$
 $$
   \begin{CD}G\times (G\times \cH_\sigma)@>>>G\times \cH_\sigma\\
   \left(g, (g',v)\right)@>>> (gg', v),
   \end{CD}
   $$
      that descends to the quotient
$$ \begin{CD}G\times E@>>>E\\
   (g, [g',v])@>>> [gg', v].
   \end{CD}
  $$

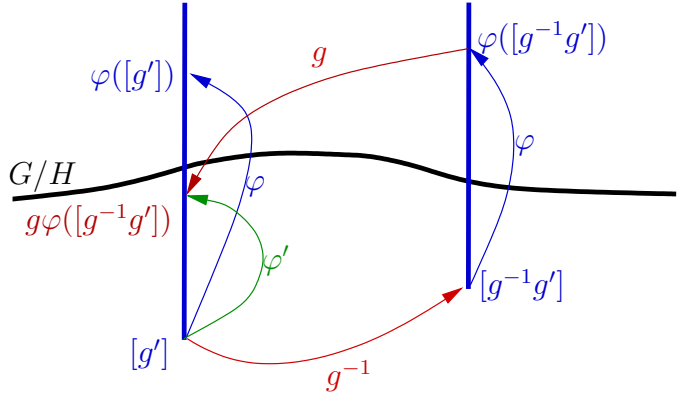
\begin{figure}  [htbp]
 \centering
 \input{bundles.tex}
 \caption{Action of $G$ on $\Gamma(E)$.}
 \label{bundles}
\end{figure}

A section of the vector bundle $E$ is a map $\varphi:G/H\rightarrow E$ such that $\pi_E\circ\varphi= \id$, where $\pi_E$ is the canonical projection $\pi_E:E\rightarrow G/H$. The space of sections of $E$, $\Gamma(E)$, carries a representation of $G$ that is the {\it representation of $G$ induced by $\sigma$}, denoted as $R_\sigma=\rInd_H^G(\sigma)$. If $\varphi$ is a section of $E$, this representation is realized as (see Fig.\ref{bundles})
  \be
   \varphi'([g']):=\left(R_\sigma(g)\varphi\right)([g'])=g\varphi([g^{-1}g']),\label{actiononsections}
 \ee
   so $\varphi'=R_\sigma(g)\varphi$ is another section.
   One can check that this is well defined on the equivalence classes.

There is another way of describing the set of sections $\Gamma(E)$. One can prove that the set of sections is in one to one correspondence with the set of equivariant functions \cite{choquet}:
\be\label{equivariant}
\cF=\{f:G\rightarrow \cH_\sigma
\; |\; f(gh)=\sigma(h)^{-1}f(g),\, g\in G,\, h\in H\}.
\ee
In fact, let $f\in\cF$. We can construct the section $\varphi$
$$\label{diagrama}\begin{CD}G/H@>\varphi>>E\\
[g]@>>>[g, f(g)],\end{CD} $$
 which is well defined due to the equivariance property. In the other direction, if we are given the section $\varphi$, then $f$ inherits its equivariance property. Unless stated otherwise, we do not impose a priori regularity conditions (such as continuity, smoothness, or square integrability) on sections, nor  on the associated equivariant functions. Such assumptions depend on the underlying framework (e.g., topological, differentiable, or Hilbert space settings) and will be implicitly imposed when needed.

From Equation (\ref{actiononsections})
one can see that the action of the group $G$ on $\cF$ is by left translation:
\be\label{9}
(R_\sigma(g)\varphi)([g'])=g\varphi([g^{-1}g'])=g[g^{-1}g', f(g^{-1}g')]= [g', f(g^{-1}g')],
\ee
and from Equation (\ref{equivariant}) we can write that
\be\label{10}
(R_\sigma(g)\varphi)([g'])= [g', f'(g')]\equiv [g',(gf)(g')].
\ee
Putting together equations (\ref{9})  and (\ref{10}) we get that
$$
(gf)(g')=f(g^{-1}g').
$$

If the representation $\sigma$ is unitary, with inner product $\langle \cdot,\cdot\rangle_{\sigma}$ in $\cH_\sigma$, then we can define an inner product on  $\cF$, so that the induced representation is also unitary. We first realize that for $f_1,{f_2}\in \cF,\, g\in G\hbox{ and } h\in H$  we have
$$
\langle f_1(gh), {f_2}(gh)\rangle_\sigma=\langle\sigma^{-1}(h) f_1(g),\sigma^{-1}(h) {f_2}(g)\rangle_\sigma=
\langle f_1(g), {f_2}(g)\rangle_\sigma ,
$$
so this quantity is constant over the coset space $gH$. Hence, it is  a function on $G/H$. We assume now that $G/H$ has an invariant measure with respect to $G$, and denote it by $\mu$.  Then, one can define an inner product on the space of sections
$$
\langle f_1, {f_2}\rangle:=\int_{G/H}\langle f_1(g), {f_2}(g)\rangle_\sigma\,\rd\mu(gH) .
$$
One can restrict to the square integrable sections. This defines the Hilbert space carrying the representation of $G$ induced from  $\sigma$.

\paragraph{About the existence of invariant measures in $G/H$.} We consider the modular functions $\Delta_G$ and $\Delta_H$, that relate the right invariant with the left invariant measures of the respective groups. If $\Delta_G\big|_H=\Delta_H$, then an invariant measure  on $G/H$ exists. This will be the case in the groups that we will consider in this paper.

There is, however, a way to construct a Hilbert space and a unitary representation if the measure on $G/H$ is only {\it quasi invariant}, that is, if the sets of null measure are invariant. Then a factor proportional to the square root of the positive {\it Radon-Nykodim derivative} appears in the transformation rule of the sections. For this and other details, the reader can consult, for example,  Ref. \cite{va}.

\hfill$\blacksquare$

\subsection{Induced representations for semidirect products}

Let $N$ be a locally compact, second-countable, abelian Lie group. It is a well known result that every unitary  irreducible representation  (UIR)  of any such group is one-dimensional.

\begin{definition} A continuous homomorphism $\chi: N\rightarrow S^1$,  with $S^1$ the set of complex numbers of modulus one is a  (unitary) character of $N$, i.e.,
 $$
 |\chi(g)|=1,\qquad \chi(gg')=\chi(g)\chi(g'),\qquad \forall g,g'\in N.
$$
 So a character is a continuous UIR of $N$.

  The {\it dual } or {\it Pontryagin dual group}  of $N$, denoted by $\widehat N$, is the set of all characters of $N$. 
  It is a group,  with group law the pointwise multiplication. Moreover $\widehat N$ is also a locally compact (in the compact-open topology), abelian group and $\widehat{\widehat N}$ is naturally isomorphic to $N$.

  \end{definition}

  \begin{example}\label{characterr2+}
  Let $N=(\R^n, +)$. Then a character in $\widehat N$ is of the form
 $$
  \chi_p (x)=\exp\ri \langle p, x\rangle,\qquad {p}\in (\R^n)^*\simeq \R^n, \, \forall\,  {x}\in \R^n,
  $$
  where $ \langle p, x\rangle=p(x)$ is the natural pairing between a vector space and its dual. So
  $\hat N=((\R^n)^*,+)\simeq (\R^n,+)$.

  \hfill$\blacksquare$
  \end{example}
   This example is going to be relevant  in what follows.

\medskip

Let $G$ be a topological group. A subgroup $H\subseteq G$ is called a \emph{closed subgroup} if $H$ is a closed subset of $G$ with respect to the topology of $G$. In this case, the quotient space $G/H$, endowed with the quotient topology, is Hausdorff. Moreover, if $G$ is a Lie group and $H$ is closed, then $H$ is an embedded Lie subgroup of $G$, and $G/H$ carries a natural smooth manifold structure such that the canonical projection $G \to G/H$ is a smooth submersion.

For groups $G$ that are a semidirect product of an Abelian normal subgroup $N$ times a closed subgroup $H$, i.e.,  $G=N\rtimes H$, the induced representation takes a more concrete form. This is Mackey's theorem, that will be stated at the end of this section. We first recall the definition of semidirect product.

\begin{definition}\label{semidirectproduct} A topological group  $G$ with a closed subgroup  $H$ and a normal subgroup $N$ is said to be a semidirect product, denoted as $G=N\rtimes H$, if the following conditions are satisfied:
\renewcommand{\theenumi}{\alph{enumi}}
\begin{enumerate}
\item[(a)]
There is  a homeomorphism $N\times H\rightarrow G$, so any element $g\in G$ can be uniquely decomposed as $(n, h)\rightarrow g=n\,h$, with  $n\in N$ and $h\in H$.

\item[(b)]
The group law is of the form
\be\label{semidirectproductf}
(n_1, h_1)(n_2,h_2)=(n_1\,h_1n_2h_1^{-1}, h_1 h_2).
\ee

\end{enumerate}
Note that in the  semidirect product, $G=N\rtimes H$,  there is a homomorphism $\tau : H\,\to\, \mathrm{Aut} (N)$  in such a way that $\tau  (h) n:=hnh^{-1}$.
\hfill$\blacksquare$
\end{definition}
The semidirect products that we will consider will be regular semidirect products. Regularity is a property on the measurability of the space of orbits and we will not enter in such discussion here. See Ref. \cite{vasusy} for more details. 

We will assume  also that the normal subgroup is Abelian. Then, we can define an action of $G$ on the dual group  $\widehat N$ as follows
\be\label{0011}
\begin{CD}
G\times \widehat N @>>>\widehat N\\
(g, \chi)@>>>g\chi
\end{CD}  \qquad \hbox{such that}\qquad (g\chi)(n)=\chi( g^{-1} ng).
\ee

\begin{definition}For every $\chi\in\widehat N$ we define
\renewcommand{\theenumi}{\alph{enumi}}
\begin{enumerate}
\item The {\it  isotropy group}:
$$
G_\chi=\{g\in G\; |\; g\chi=\chi\}.
$$

\item The {\it orbit}:
$$
\cO_\chi=\{g\chi\; \hbox{ for all } g\in G\}\simeq G/G_\chi.
$$
\item The {\it little group}:
$$
H_\chi=H\cap G_\chi.
$$
\end{enumerate}

\hfill$\blacksquare$
\end{definition}
\hyphenation{se-cond}
We then have:
\begin{theorem} \label{mackey} {\rm (Mackey)\cite{ma1,ma2}}. Let  $G=N\rtimes H$ be a locally compact, second countable group that is a regular semidirect product, with $N$ an abelian,  normal subgroup and with $H$ a closed subgroup. Let $\chi\in\hat{N}$ and let $H_\chi$ be the little group of $\chi$. Let $\sigma:H_\chi\rightarrow   \rAut (\cH_\sigma)$ a unitary representation of  $H_\chi$  on the Hilbert space $\cH_\sigma$. Then, there is a representation of the isotropy group $G_\chi$  on $\cH_\sigma$, denoted as $\chi\sigma$, given for  $g=n\,h\in G_\chi$ as
$$\chi\sigma (n\,h)=\chi(n)\sigma(h).
$$
Moreover, the representation $\rInd_{G_\chi}^G(\chi\sigma)$, (denoted usually as $\rInd_{H_\chi}^G(\chi\sigma)$ for semidirect products) is a unitary irreducible representation of the group $G$ on the space of square integrable sections $\cF\simeq\Gamma(E)$, where $E=(G\times \cH_\sigma)/G_\chi$.

The representations computed in this way, corresponding to characters in the same orbit, are all equivalent and are inequivalent if they belong to different orbits.

All the unitary irreducible representations of $G$ can be obtained, up to equivalence, in this way.

\hfill$\blacksquare$

\end{theorem}

\section{The Euclidean group $\rE(2)$}\label{euclidean-sec}

\label{sec:E(2)}

We consider the group of motions in the Euclidean plane $\rE(2)\equiv\rISO(2)$  (where `I' stands for `inhomogeneous'), that is, translations and rotations in the plane. We will denote an element of this group as
$
({\mathbf a}, R_\theta),
$
where
$$
{\mathbf a}=\begin{pmatrix} a_1\\a_2\end{pmatrix}, \;\; a_i\in\R\,;\qquad
R_\theta=\begin{pmatrix}\cos \theta&- \sin\theta\\\sin\theta&\cos\theta\end{pmatrix}, \qquad \theta\in[0,2\pi ).
$$
The action of $\rE(2)$  on the 2D Euclidean space, $\rE(2)\times \R^2\,\to\, \R^2$, is
$$
{\mathbf y}=R_\theta\, {\vx}+{\mathbf a},
$$
which gives the group law
\be
({\mathbf a},R_\theta)\cdot ({\mathbf b}, R_\phi)=({\mathbf a}+R_\theta\,{\mathbf b}, R_\theta\,R_\phi )=
({\mathbf a}+R_\theta\,{\mathbf b},R_{\theta+\phi}),
\label{grouplawE(2)}\ee
with the identity
$\rId=(\mathbf 0, \id)$ and the inverse element of $({\mathbf a}, R_\theta)$
\be\label{inverse}
( {\mathbf a}, R_\theta,)^{-1}=(-R_\theta^{-1}{\mathbf a},R_\theta^{-1}) .
\ee
Let us  denote  pure translations as
$\mathcal{T}_{\mathbf{a}}:=(\mathbf{a}, \id)$ and pure rotations as $\mathcal{R}_\theta:= (\mathbf{0}, R_\theta)$.
The group E(2) is a semidirect product, i.e.   $\rE(2)=\cT_2\rtimes \rSO(2)$, where $\cT_2=\R^2$ is a normal, abelian subgroup:
$$
({\mathbf a}, R_\theta)\,\cT_{\mathbf b}\,({\mathbf a},R_\theta)^{-1}=\cT_{ R_\theta\,{\mathbf b}}.
$$
and $\rSO(2)$ is a closed  subgroup.
The condition (a)  in the Definition \ref{semidirectproduct} is satisfied since
$$ \label{0024}
( {\mathbf a},R_\theta)= ( {\mathbf a},\id) ( {\mathbf 0},R_\theta)=
\cT_{\mathbf a}\,\cR_\theta,
$$
and this decomposition is unique.
Similarly  for condition  (b) in the same definition, since
$$ \cT_{\mathbf a}\cR_\theta\cT_{\mathbf b}\cR_\theta^{-1}=\cT_{{\mathbf a}
+\cR_\theta{\mathbf b}}, $$
and we easily recover the group law (\ref{grouplawE(2)}).

\medskip

The left invariant vector fields on $\rE(2)$ are
\begin{align*} 
 Q_1&:=\rd L_g|_e\,\partial_{{g'}^1}=\cos\theta\, \partial_{a_1}-\sin \theta\, \partial_{a_2},\nonumber\\[0.3cm]
 Q_{2}&:=\rd L_g|_e\,\partial_{{g'}^1}=\sin\theta\,\partial_{a_1}+\cos\theta\,\partial_{a_2},\nonumber\\[0.30cm]
 J&:=\rd L_g|_e\,\partial_{{g'}^0}=\partial_\theta,
 \end{align*}
 with commutation rules
 \be \label{crE(2)}
 [J, Q_{1}]=-Q_{2},\qquad [J, Q_{2}]=Q_{1},\qquad [Q_{1}, Q_{2}]=0 .
\ee
 It is immediate to see that the element in the enveloping algebra of $\rE(2)$
  $$ 
 {\mathbf Q}^2=(Q_{1})^2+(Q_2)^2=\partial_{a_1}^2+\partial_{a_2}^2
 $$
  is a quadratic Casimir.

\medskip 

Since we are going to deal with spinors, the group $\rSO(2)$ will be substituted by its (double cover) Spin group, 
$$\rSpin(2)= \left\{\begin{pmatrix}a&-d\\d&a\end{pmatrix}\;\big|\,\; a,d\in\R,\quad  a^2+d^2=1\right\}.$$  It is computed in Appendix \ref{spingroupE(2)}. It is isomorphic to $\rSO(2)$ itself, but there is a double cover homomorphism 
$$\begin{CD}\rSpin(2)@>\varphi>>\rSO(2)\\[0.2cm]
\begin{pmatrix}a&-d\\d&a\end{pmatrix}@>>>\begin{pmatrix}a^2-d^2&-2ad\\2ad&a^2-d^2\end{pmatrix}.
\end{CD}$$It is convenient to  parametrize the Spin group as

$$a=\cos\frac{\theta}{2},\qquad d=\sin\frac{\theta}{2},\qquad \theta=[0,4\pi[,$$  and then,  the double cover homomorphism is given by

\be\begin{CD}\begin{pmatrix}\cos {\theta}/{2}&-\sin{\theta}/{2}\\\sin{\theta}/{2}&\cos{\theta}/{2}\end{pmatrix}@>\varphi>>\begin{pmatrix}\cos \theta&-\sin\theta\\\sin\theta&\cos\theta\end{pmatrix}.\end{CD}\label{homE(2)}\ee

\medskip

We consider now the subgroup of $\rSL(2,\C)$  denoted as $\widetilde{\rE}(2)$ and  consisting of the 
  set of matrices
 \be\left\{\begin{pmatrix}z&z^{-1}a\\0&z^{-1}\end{pmatrix}, \quad |z|=1,\quad a\in \C\right\}.\label{smallgroupE(2)}\ee It forms a group under matrix multiplication:
$$\begin{pmatrix}z&z^{-1}a\\0&z^{-1}\end{pmatrix}
\begin{pmatrix}z'&{z'}^{-1}a'\\0&{z'}^{-1}\end{pmatrix}=
\begin{pmatrix}zz'&(z{z'})^{-1}(a+z^2 a')\\0&(z{z'})^{-1}\end{pmatrix},$$ with inverse
$$\begin{pmatrix}z^{-1}&z(-z^{-2}a)\\0&z\end{pmatrix}.$$
 We denote an element of the group as $\tilde g=(a, z)$. 
 The subgroup of elements $(b, 1)$ is a normal subgroup, since
  $$(a,z)(b,1)(a,z)^{-1}=(z^2b, 1).$$ According to Definition~\eqref{semidirectproduct},  $\widetilde{\rE}(2)$ has the structure of a semidirect product  ($ \widetilde{\rE}(2)=N\rtimes H$), with $N=\C\cong\R^2$ that can be identified as the translations, and $H$ as $\rSpin(2)$ (the double cover of $\rSO(2)$), due to its action on  $N$:
$$\begin{CD}N@>\tau(z)>>N\\
a@>>>z^2a,\end{CD}$$ 
Notice that the map $\widetilde{\rE}(2)\rightarrow \rE(2)$ is two to one, so $\widetilde{\rE}(2)$ is a double cover of ${\rE}(2)$,  i.e.
\be\label{0046}
\begin{CD}\widetilde \rE(2)@>>> \rE(2)\\
\tilde g=(a,z)@>>> g=({\mathbf a}, R_{\theta}),\end{CD}
\ee
where $a=a^1+\ri a^2\in \C$, ${\bf a}=(a^1,a^2)\in \R^2$ and $z=\re^{\ri\theta/2},\, \theta \in [0,4\pi)$. 

This group and its representations appear naturally in the context of the null mass representations of 
the Poincar\'{e} group in $D=4$. In fact, the massless orbits of $\rSO(1,3)$ on the momentum space are characterized by a momentum $\mathbf{p}_w=(w,0,0,w)$, which, in terms of the Pauli matrices
$$X_\mathbf{p}=\sigma^\mu p_\mu =\begin{pmatrix}p_0+p_3&p_1-\ri p_2\\
p_1+\ri p_2&p_0-p_3
\end{pmatrix},$$ is just
$$X_w=\begin{pmatrix}w&0\\0&0\end{pmatrix},\qquad w\neq 0.$$ Then it is easy to show that the group elements $A\in\rSL(2,\C)$ leaving stable this momentum, i.e.   $ X'_w=AX_wA^\dagger=X_w$, are

 $$A=\begin{pmatrix}z&\tilde a\\0&z^{-1}\end{pmatrix},\qquad  |z|=1,\; 
 \;  \tilde a\in\C.  
$$
 Redefining $\tilde a=z^{-1} a$ we obtain the expression (\ref{smallgroupE(2)}). As far as we know, this is the easiest way to show that the Euclidean group $\rE(2)$ is the little group of the massless orbits of $\rSO(1,3)$ \cite{va}.
 
 In what follows, we will write $g$ for the elements $\tilde g \in  \widetilde{\rE}(2)$
whenever no confusion is likely to arise

\section{ Unitary, irreducible representations of $\rE(2)$}\label{sec:UIR(2)}

 { We will proceed by identifying each piece in Mackey's Theorem \ref{mackey}. }
 
\subsection{Orbits and equivariant functions}

First, we will  obtain an explicit characterization of the equivariant functions  in (\ref{equivariant}). We recall from Example \eqref{characterr2+} that a character of $\cT_2=(\R^2,+)$ is of the form
  \be 
  \chi_{\mathbf q}({{\bf a}})
 =\re^{\ri\langle{\mathbf q},{\bf a}\rangle},\qquad {{\bf a} }\in \cT_2,\qquad {\mathbf q}\in \widehat \cT_2 ,\nonumber
\ee
where $\langle{\mathbf q},{\bf a}\rangle={\mathbf q}\cdot {\bf a}=q_1\, a^1+q_2\, a^2$.
In complex language, $q=q_1 +\ri q_2$, $a=a^1 +\ri a^2$ and $\langle{\mathbf q},{\bf a}\rangle=\mathrm{Re}(\bar q a)$.

The action of an element $g=(a, z)\in \widetilde{\rE}(2)$ on a character $\chi_{\mathbf q}$ is defined, according to \eqref{0011}, as
\be
(g\chi_{\mathbf q})({b},\id):=\chi_{\mathbf q}\left(g^{-1}({b},\id)g\right)= \chi_{\mathbf q}(R^{-1}_\theta{\bf b}, \id).\nonumber
\ee

We identify $\R^2$ with its dual by means of the Euclidean metric and, being $R_\theta$ orthogonal, we obtain  that 
$g\chi_{\mathbf q}\equiv \chi_{\mathbf q'}$ with
 \be 
 {\vq'}=R_\theta\,{\vq} \qquad \hbox{or, in complex notation,}\qquad  q'=z^2 q.\nonumber
  \ee
The translations, $(a,\id)$, act trivially on $\vq$.

With respect to this action, we have two kinds of orbits in the momentum space  $\widehat \cT_2$:
   
{\bf 1) The origin.} ${\vq} =0$, which is a non regular, single point orbit, whose
  isotropy group 
   is the full $\widetilde{\rE}(2)$. The little group of ${\vq} =\mathbf 0$ 
  is $\rSpin(2)$.  The base manifold
  is just a point and   the fiber bundle $E$ is the Hilbert space $\cH_\sigma$ supporting a UIR of the little group. The UIRs of $\rSpin(2)$ are one-dimensional and labelled  by integers $m,\; m\in \Z$, with $\cH_m=\C$ and
  $$
   \rSpin(2)\ni \, z=\re^{\ri\theta/2}\;\xrightarrow[ m]{\rm UIR} \;\re^{im\theta/2}.$$
  The character $\chi_0$ is 1, so the translations are represented trivially. 
   Therefore, according to Theorem \eqref{mackey}, we have
 \be \label{44}
 \rInd_{{\rSpin}(2)}^{\widetilde{\rE}(2)}(m)({\bf a}, \re^{\ri\theta/2})=\re^{\ri m\theta/2} ,\qquad  \theta\in [0,4\pi[.\ee
  There are two kinds of representations:
   \begin{itemize}
   \item $m$ {\bf even}. So $m =2\,n$, hence  $\re^{\ri m\theta/2}=\re^{\ri n\theta}$.  Since $\re^{\ri n2\pi}=1$, we have periodicity in the interval $[0, 2\pi[$, thus the variable $\theta$ can be taken in that interval. These are also representations of $\rE(2)$.

   \item  $m$ {\bf odd}. Now $m=2n+1$, $\re^{\ri (2n+1)2\pi/2}=-1$ and the identity is recovered only after a turn of $4\pi$. They  appear only for the double cover.
\end{itemize}

  {\bf 2) The regular orbits.} ${\vq} \neq 0$, which are circles of radius $q=|{\vq} |>0$, centered in the origin $\mathbf 0$.  Fixed a point of the orbit,
       ${\mathbf q}_l$,
          its  isotropy group, $G_{{\mathbf q}_l}$, is $\cT_2\rtimes\{\pm\id\}$ and its  little group $H_{{\mathbf q}_l}=\{\pm\id\}$.  The base manifold is
      $$
      G/G_{{\mathbf q}_l}=\widetilde{{\rE}}(2)/\cT_2\rtimes\{\pm\id\}={\rSO}(2),
      $$
and the bundle  $E$  of \eqref{bundle}  is a (trivial) complex line bundle over the circle ${\rSO}(2)\simeq S^1$, with fiber the Hilbert space $\cH_\sigma=\C$.

 The representations are characterized by a vector ${{\mathbf q}_l}\neq 0$ in the orbit and we denote them by $U_{{\mathbf q}_l}(\tilde g)$. They are  infinite dimensional. To see this, we decompose an element of 
  $\widetilde{\rE}(2)$ as
\be
({\bf a}, z)=({\bf a}, 1)(0,z)=(0,z)(0,z)^{-1}({\bf a}, 1)(0,z)=(0,z)(z^{-2}{\bf a}, 1),\nonumber
\ee
so an equivariant function \eqref{equivariant} is of the form
\be\label{equivariantfunction}
f({\bf a}, z)=\re^{-\ri {{\mathbf q}_l}\cdot  R_\theta^{-1}\,{\mathbf a}}f(0,z)=\re^{-\ri R_\theta {{\mathbf q}_l}\cdot{\mathbf a}} f(0,z)
=: \re^{-\ri R_\theta {{\mathbf q}_l}\cdot{\mathbf a}}\tilde f(z),
\ee
with $z=\re^{\ri \theta/2}$. On the other hand, there are two irreducible representations for $\{\id, -\id\}$ on $\C$. One is the trivial one, where both elements go to 1 and the other takes $\pm \id\rightarrow \pm 1$. 
 Imposing those conditions on the equivariant functions we obtain the two possibilities
\be
\tilde f(-z)=\pm \tilde f(z).\label{parity}
\ee
The action of an element  $ g=({\bf b}, u=\re^{\ri\phi/2})$ of $\widetilde {\rE}(2)$  
on the equivariant functions is
\[
f^{ g}({\bf a},z)=\big(({\bf b}, u)f\big)({\bf a},z)=f\big(({\bf b}, u)^{-1}({\bf a},z)\big)=\re^{-\ri R_{\theta}\mathbf{q}_l\cdot(\mathbf{a}\mathbf{-b})}\tilde f(u^{-1}z).
\]
Since  $f^{ g}({\bf a},z)$ is an equivariant function,  it has the form (\ref{equivariantfunction}), and then,  the action of $\widetilde{\rE}(2)$ on 
$\tilde f(z)$ is
$$
{\tilde f}^{ g}(z)=\big(U_{{\mathbf q}_l}( g)\, \tilde f \big) (z)=\big(({\bf b}, u)\tilde f\big)(z)=\re^{\ri R_\theta {\mathbf q}_l\cdot {\mathbf b}}\tilde f (u^{-1}z).
\nonumber$$
By simply denoting $\tilde f(\theta)$ instead of $\tilde f(z)$, we rewrite ${\tilde f}^{ g}(z)$ as
\be\label{tequivariantsections}
{\tilde f}^{ g}(\theta)=\left(U_{{\mathbf q}_l}( g) \,\tilde f \right)  (\theta)=\big(({\bf b}, \re^{\ri\phi/2}) \tilde f\big)(\theta)=\re^{\ri R_{\theta}\mathbf{q}_l\cdot\mathbf{b}}\tilde f(\theta-\phi).
\ee

\subsection{Lie algebra representation}\label{lie algebra rep e(2)-sec}

We now  compute the corresponding representation of the Lie algebra of $\widetilde \rE(2)$, ${\mathfrak e}(2)$.
 We consider an infinitesimal transformation
  \be
 U(\delta{{\bf b}},{\delta\phi})=
 e^{\ri\,\delta{{\bf b}}\cdot {\widehat\vcq}}\,e^{-\ri\,\delta{\phi}\,\widehat J}
 =\rId+\ri {\delta{\mathbf b}}\cdot\widehat\vcq-\ri \delta\phi \widehat J+\cO(\delta {\bf b}, \delta\phi)^2
  \label{infinitesimaltransformation}\ee
   were $\delta{\mathbf b}=(\delta{b_1},\delta{b_2})$ and $ \delta \phi$ are the  infinitesimal parameters
 and $-\ri\widehat\vcq$ and $-\ri \widehat J$  are the generators of $\tilde{\mathfrak e}(2)$.  
 Written in this way,  $\widehat\vcq$ and $\widehat J$ are Hermitian operators, as it is customary in physics.
  From (\ref{tequivariantsections}) we can develop in power series the function ${\tilde f}^{\tilde g}(\theta)$ around the identity, i.e.,  $\phi=0,\,{\bf b}=0$
\be\begin{array}{lll}
  {\tilde f}^{\tilde g}(\theta)&=&\ds \left[\tilde f^{\tilde g}(\theta)
  +\frac {\partial {\tilde f}^{\tilde g}(\theta)}{\partial b_i}\,\delta b_i
  +\frac {\partial {\tilde f}^{\tilde g}(\theta)}{\partial \phi}\,\delta\phi\right]_{\substack{\mathbf b=\mathbf 0\\\phi=0}}
  +\cO(\delta{{\bf b}}, \delta\phi)^2\\[0.3cm]
&=& \ds \tilde f(\theta)+\ri R_\theta \vq_l\cdot\delta{\mathbf b}\,\tilde f(\theta)-\frac{\partial \tilde f (\theta)} {\partial\theta} \delta\phi  +\cO(\delta{{\bf b}}, \delta\phi)^2. \label{powerseriesequivariant}
\end{array}\ee
On the other hand,  under the action of the  infinitesimal transformation  $U(\delta{b},{\delta\phi})$ \eqref{infinitesimaltransformation} we get
\be\begin{array}{lll}
  {\tilde f}^{\tilde g}(\theta)&=&\ds \left({ U}(\delta{{\bf b}},{\delta\phi}) {\tilde f}\right) (\theta)\\[0.3cm]
 &=&\ds \left(\rId+\ri {\delta{\mathbf b}}\cdot\widehat\vcq-\ri \delta\phi \,\widehat J+\cO(\delta{ {\bf b}}, \delta\phi)^2\right)
  \tilde f(\theta)\\[0.3cm]
 &=& \tilde f(\theta)+\ri  {\delta{\mathbf b}}\cdot\widehat\vcq\, \tilde f(\theta) -\ri \delta\phi \,\widehat J \tilde f(\theta)
    +\cO(\delta{\mathbf b}, \delta\phi)^2 \tilde f(\theta). \label{orderoneaction}
\end{array}\ee
Comparing expression \eqref{powerseriesequivariant} and  \eqref{orderoneaction}, up to the first order in the infinitesimal parameters, we obtain the explicit expression of the operators  $\widehat\vcq=(\widehat Q_{1},\widehat Q_2)$ and $\widehat J$ in this representation

 \begin{align}
 \widehat Q_{1} &=\ds q_1\cos\theta -q_2\sin\theta\, , \nonumber\\[0.3cm]
   \widehat Q_{2}&=\ds q_1\sin\theta +q_2\cos\theta, \nonumber\\[0.3cm]
 \widehat J&=\ds-\ri\,\frac{\partial } {\partial\theta},\nonumber
 \end{align}
 were ${\bf q}_l^t=(q_1,q_2)$. One can check that the commutation relations are
 \[
 [\widehat  J,\widehat Q_1]=\ri \widehat Q_2,\quad [\widehat J,\widehat Q_2]=-\ri \widehat Q_1 ,\quad 
 [\widehat Q_1, \widehat Q_2]=0.
 \]
 If one multiplies by $\ri$ these generators  (here the operators are Hermitian), one obtains the commutation relations in
 \eqref{crE(2)}.
 Moreover,
\[
\mathcal{C}^2\tilde f(\theta)=\widehat \vcq^2\tilde f(\theta)=\left((\widehat Q_1)^2+(\widehat Q_2)^2\right)\tilde f(\theta)=\left(q_1^2+q_2^2\right)\tilde f(\theta),
\]
 and $q_1^2+q_2^2$ is the radius squared, constant on the orbit. This means that the functions $\tilde f(\theta)$ are eigenvectors  of 
 $\widehat \vcq^2$ with eigenvalue $q_1^2+q_2^2$.


 \subsection{Basis of eigenvectors of $\hat J$}\label{eigenvaluesj}

 We look now for the eigenvectors of the operator $J=-\ri\partial_\theta$. Hence, we have to find functions
 $\tilde  f (\theta)$  such that:
\be
-\ri\, \frac{\partial \tilde f (\theta)} {\partial \theta}=\lambda\tilde f(\theta) \implies
 \tilde f(\theta)=A\,\re^{+\ri \lambda\theta},\qquad A,\lambda\in \C.
 \nonumber\ee Imposing the boundary conditions,
\be
\tilde f (\theta)=\tilde f (\theta+ 4\pi)\implies \lambda=  \frac m 2,\quad m\in \Z.\nonumber
\ee
So the eigenfunctions of $J$, with eigenvalue $m\in \Z$ are the functions
 \be\label{00043}
 \tilde{f}_m(\theta)=A_{m}\,\re^{\ri m\theta/2}.
 \ee
 We have to impose also the parity conditions (\ref{parity}) on $ \tilde{f}_m(\theta)$ in \eqref{00043}. Recalling that $z=\re^{\ri\theta/2}$ and, obviously, $-z= \re^{(\ri\theta+2\pi)/2}$, we can distinguish between  $m$ even or odd, i.e.
$$\begin{array}{lll}
 \tilde{f}_{2n}(\theta)=A_{2n}\, \re^{\ri n\theta}, \qquad &\rm{with}\;&\tilde f_{2n}(\theta+2\pi)=f_{2n}(\theta), \\[0.3cm]
 \tilde{f}_{2n+1}(\theta)=A_{2n+1}\, \re^{\ri (2n+1)\theta/2}, \qquad &\rm{with}\;\; &\tilde f_{2n+1}(\theta+2\pi)=-f_{2n+1}(\theta).
 \end{array}$$
 The functions $\tilde{f}_{2n}(\theta)$ are indeed periodic with period $2\pi$, while for the $\tilde{f}_{2n+1}(\theta)$ one has to take a second turn to go back to the original value. We have then two different representations, the `bosonic' one, spanned by $\{f_{2n}(\theta)\}_{n=0}^\infty$ and a `fermionic' one, spanned  by $\{f_{2n+1}(\theta)\}_{n=0}^\infty$.

There is an invariant inner product with respect to which the basis vectors (eigenvectors) of each representation are orthogonal:
\begin{align}\langle\tilde f_{2n}, \tilde f_{2n'}\rangle&=\frac 1{2\pi}\int_0^{2\pi}\tilde{f}^*_{2n}(\theta)\tilde{f}_{2n'}(\theta)\rd \theta=
|A_{2n}|^2\delta _{nn'},\nonumber\\[0.3cm]\label{60}
\langle\tilde f_{2n+1}, \tilde f_{2n'+1}\rangle&=\frac 1{4\pi}\int_0^{4\pi}\tilde{f}^*_{2n+1}(\theta)\tilde{f}_{2n'+1}(\theta)\rd \theta=
|A_{2n+1}|^2\delta _{nn'}.\nonumber
\end{align}

 Choosing $A_{2n}=A_{2n+1}=1$ we end up with the orthonormal bases of the representation spaces, formed by eigenvectors of $J$ 
 \eqref{00043}.

We want now to compute the matrix  elements of the infinite matrix representations. 

Let $\tilde g=({\bf b}, u)=({\bf b}, \re^{\ri\phi/2})\in \widetilde{\rE}(2)$ \eqref{0046}. Its action on the equivariant sections is given by (\ref{tequivariantsections}), so
\[\begin{array}{lll}
 U_{\mathbf{q}_l}({\bf b}, \re^{\ri\phi/2})_{m'm}&=&\langle \re^{\ri m'\theta/2},  U_{\mathbf{q}_l}({\bf b}, \re^{\ri\phi/2}) \re^{\ri m\theta/2}\rangle\\[10pt]
&=&\ds\frac 1{4\pi}\int_0^{4\pi}\re^{-\ri m'\theta/2}\re^{\ri R_\theta\mathbf{q}_l\cdot \mathbf{b}}\re^{\ri m(\theta-\phi)/2}\rd\theta.
\end{array}\]
Note that $R_\theta\mathbf{q}_l\cdot \mathbf{b}$ can be expressed as 
 $ A\cos\theta+B\sin\theta$, 
where $A=q_1b_1+q_2b_2$ and $B=q_1b_2-q_2 b_1$. Equivalently, we can write  this quantity as 
\[
R_\theta\mathbf{q}_l\cdot \mathbf{b}=R\cos(\theta-\theta_0),\quad\rm{with}\qquad A=R\cos\theta_0,\quad B=R\sin\theta_0.
\]
Hence
\[
 U({\bf b}, \re^{\ri\phi/2})_{m'm}=\frac{\re^{-\ri m\phi/2}}{4\pi}\int_0^{4\pi}\re^{-\ri m'\theta/2}
 \re^{\ri R\cos(\theta-\theta_0)}\re^{\ri m\theta/2}\rd\theta.
 \] 
 By a change of variable $\psi=\theta-\theta_0$  we obtain
\[
U({\bf b}, \re^{\ri\phi/2})_{m'm}=\frac 1{4\pi}\re^{-\ri m\phi/2}\re^{\ri(m-m')\theta_0/2}\,
\int_0^{4\pi}\re^{\ri(m-m')\psi/2}\re^{\ri R\cos\psi}\rd \psi,
\] and since $m,m'$ are both even or both odd, $m-m'=2n$ and we can just integrate over the interval $[0,2\pi[$. Comparing with the Hansen-Bessel    formula  (see, for example, \cite{watson1958}, Section 2.2, (5)) for the  Bessel functions  of first kind:
\be
J_n(z)=\frac {(-\ri)^n}{2\pi}\int_0^{2\pi} \re^{\ri z\cos t}\re^{\ri nt}\rd t,\nonumber
\ee
we finally obtain
\[
 U({\bf b}, \re^{\ri\phi/2})_{m'm}=\re^{-\ri m\phi/2}\,\re^{\ri (m-m')(\theta_0+\pi/2)/2}J_{(m-m')/2}(R).
 \] 
 One has that $R=|\mathbf{q}|\cdot|\mathbf{b}|$, and choosing a representative of the orbit $\mathbf{q}=(|q|,0)$, we get that $\tan\theta_0=b_2/b_1$. Then 
one can compare with, for example, the expression  in  \cite{tung} (Theorem 9.4), where, however, a different normalization is chosen for the functions $\tilde f_n(\theta)$.

\section{The Poincar\'{e} group $\rP(1,1)$} 

\label{sec:Poincare}
The 2D Poincar\'{e} group is the group of motions of the 2D Minkowski space, whose  metric is
$$\eta_{\mu\nu}=\diag(1,-1),\qquad \mu,\nu=0,1.$$

The group that leaves invariant the metric is the pseudo-orthogonal group $\rO(1,1)$. We will restrict our analysis to 
$ \rSO_0(1,1)$, its connected component of the identity, that  can be parametrized  by a real parameter $\chi$ as
$$
\Lambda_\chi=\begin{pmatrix}\cosh\chi&\sinh\chi\\\sinh\chi&\cosh\chi\end{pmatrix},\qquad \chi\in \R.
$$  
In fact, $\rSO_0(1,1)\cong \R\cong\R^\times_+$ via the  group isomorphisms
$$
\chi {\, \in (\R,+)} \longleftrightarrow { \re^\chi \in (\R^\times_+,\cdot)} \longleftrightarrow \Lambda_\chi\in \rSO_0(1,1).
$$
 Note $\R^\times_+\subset \R^\times$ is the multiplicative group of the positive real numbers. 
 
 In the Appendix \ref{spingroupP(1,1)}, we compute its  Spin group, $\rSpin(1,1)\cong\R^\times$, that can be parametrized as 
$$\rSpin(1,1)=\{\pm\re^{\chi/2}\;|\; \chi\in \R\},$$ and there is a double cover homomorphism to $\rSO_0(1,1)\cong\R^\times_+\cong \R$
$$
\begin{CD}\rSpin(1,1)@>>>\rSO_0(1,1)\\
\pm\re^{\chi/2}@>>>\re^{\chi}.\end{CD}
$$

Together with the translations 
$\mathbf a\in \R^2$,  it acts on the two dimensional Minkowski space as
$$
{\mathbf y}=\Lambda_\chi\, {\vx}+{\mathbf a}.
$$
 We will denote this group as $\rP(1,1)\equiv \rISO_0(1,1)$. The structure of this group is  similar to the structure of the Euclidean group, so it is also a semidirect product $\rP(1,1)=\cT_2\rtimes \rSO_0(1,1)$. The group law and the inverse can be read in (\ref{grouplawE(2)}) and (\ref{inverse}), substituting $R_\theta$ by $\Lambda_\chi$. 

 We will also denote by $\widetilde{\rP}(1,1)$ the corresponding double cover of $\rP(1,1)$.

\medskip

As in Section \ref{sec:E(2)},  the left invariant vector fields are:

\begin{align*} 
 Q_0&=\cosh\chi\, \partial_{a_0}+\sinh\chi\, \partial_{a_1},\nonumber\\[0.3cm]
 Q_{1}&=\sinh\chi\,\partial_{a_0}+\cosh\chi\,\partial_{a_1},\nonumber\\[0.30cm]
 K&=\partial_\chi,
 \end{align*}
 with commutation rules
 
\be \label{crP(1,1)}
 [K, Q_{0}]=Q_{1},\qquad [K, Q_{1}]=Q_{0},\qquad [Q_{0}, Q_{1}]=0.
\ee
It also has a quadratic Casimir in the enveloping algebra:
$${\mathbf Q}^2=Q_0^2-Q_1^2=\partial^2_{a_0}-\partial_{a_1}^2.$$

\section{The UIR of the Poincar\'{e} group in $D=2$.}\label{sec:uir(1,1)}

In this section we present  the construction the UIR of $\rP(1,1)$, following a process similar to what we do for $\rE(2)$, (see Section \ref{sec:UIR(2)}) according to Mackey's induction method.

\subsection{Orbits and equivariant functions}\label{orbitsso11}

The orbits on the momentum space are of the following types:
\medskip

\renewcommand{\theenumi}{\alph{enumi}}

  {\bf 1) The origin.} $\mathbf{q}=0$.\label{zeroorbit} The isometry group is the full $\widetilde{\rISO}(1,1)$. The base space is just a point. The little group is the full $\rSpin(1,1)$.  The UIR of $\R^\times$ are its characters, since it is an abelian group:
    \be R_{0,\lambda}(\pm\re^{\phi/2} )=\re^{\ri\lambda\phi},\qquad R_{1,\lambda}(\pm\re^{\phi/2} )=\pm \re^{\ri\lambda\phi},\quad \lambda\in\R,\label{UIR(1,1)}\ee where the factor 1/2 has been reabsorbed in $\lambda$. The character of $(\R^2,+)$ associated with $\mathbf{q}=\mathbf{0}$ is the identity, so the induced representations have the form above (\ref{UIR(1,1)}).
  \medskip  
    
 {\bf 2) Light-like orbits.}  $\mathbf{q}^2=0, \mathbf{q}\neq 0$.\label{masslessorbit} There are four  orbits, corresponding to the four disconnected pieces of the light cone minus the origin. For $\mathbf{q}=( q_0,q_1)$ one has the four combinations:
\[
 q_0=\pm q_1,\qquad\hbox{and}\qquad q_0<0 \;\hbox{ or } \;q_0>0,
\]     
with representatives ($q>0$):
        \be  \label{fourorbits}
         {{\mathbf q}_l}= \begin{pmatrix}q\\q\end{pmatrix},\qquad {{\mathbf q}_l}= \begin{pmatrix}q\\-q\end{pmatrix}, \qquad{{\mathbf q}_l}=  \begin{pmatrix}-q\\q\end{pmatrix},\qquad {{\mathbf q}_l}= \begin{pmatrix}-q\\-q\end{pmatrix}. 
        \ee 
        They are related by parity or temporal inversion, or both. In all the cases the little group is $\Z_2=\{ \pm \id\}$, the base manifold is $\rSO_0(1,1)\cong\R$ and the bundle $E=\R\times \C$.
        
        For simplicity, we define $u=\re^{\phi/2}$. Then,  following the same reasoning as for the Euclidean group (\ref{equivariantfunction}), the equivariant functions are of the form
       \be\label{equivariant1,1}
       \begin{CD}\widetilde{\rP}(1,1)@> f>>\C\\
       ({\bf a},\pm u)@>>>\re^{-i\Lambda_\phi{{\mathbf q}_l}\cdot{\bf a}} \tilde{f}(\pm u).
       \end{CD}
       \ee 
       Notice that we are writing the Spin group as
$$
\mathrm{Spin}(1,1)=\{\eta\,u\; |\; \eta=\pm1,\, u>0\},
$$ meaning a parametrization of the two sheets as 
$$\mathrm{Spin}(1,1)=\{(\eta,u)\;|\; \eta=\pm1,\, u>0\}.
$$
Then, equivariance with respect to the non-trivial element of the kernel
$\Z_2=\{\pm1\}$
implies that the two sheets of the covering are related by
$$
\tilde f(-1,u)=\varepsilon\,\tilde f(+1,u),
\qquad \varepsilon=\pm1.
$$
This condition does not impose any restriction on the continuous variable
\(u\in\R^+\); it merely identifies the values of $\tilde f$ on the
two sheets of the spin covering, according to the chosen one-dimensional
representation of the little group. We will call these representations `bosonic' ($\varepsilon =1$) and `fermionic' ($\varepsilon=-1$).

       The action of an element of the group $g=({\bf b}, \eta' v)\in \widetilde{P}(1,1)$, with $v=\re^{\zeta /2}$ on the sections $\tilde f$ is, as in (\ref{tequivariantsections}):
       \be
       \tilde f^{ g}(\eta u)=\left(U_{{\bf q}_l}({\bf b}, \eta' v)f\right)(\eta u)=\re^{\ri \Lambda_\phi\,{{\mathbf q}_l}\cdot {\bf b}}\tilde{f}(\eta\eta' v^{-1}u).\label{transformedsection}\ee
       One can be more explicit. To avoid clumsy notation, we consider bosonic representations only, and denote $\tilde f(u)$ as $\tilde f(\phi)$, since $u=e^{\phi/2}$. Let us analyze separately the four types of representations associated to the four types of orbits (\ref{fourorbits}). The action of a Lorentz transformation on $\mathbf q$ is as follows:
       $$
\Lambda_\phi\begin{pmatrix}\pm q\\\pm q\end{pmatrix}
  = \re^\phi\begin{pmatrix}\pm q\\\pm q\end{pmatrix},\qquad
\Lambda_\phi\begin{pmatrix}\pm q\\\mp q\end{pmatrix}
  = \re^{-\phi}\begin{pmatrix}\pm q\\\mp q\end{pmatrix},
$$
and defining $b^\pm=b^0\pm b^1$, the action   (\ref{transformedsection}) becomes, respectively,
\be\label{00200}
\tilde{f}^{ g}(\phi)=\re^{\pm\ri\re^\phi qb_-}\tilde f(\phi-\zeta),\qquad  \tilde{f}^{ g}(\phi)=\re^{\pm\ri\re^{-\phi} qb_+}\tilde f(\phi-\zeta).
\ee 
So only $b_+$ or $b_-$ act non trivially.

The mass shell condition (orbit constraint) is
$$
q^2=q_0^2- q_1^2=(q^0q_0+q_1)(q_0-q_1)=0,
$$ 
so the solutions can be  chiral, $q_0=q_1$, or antichiral,  $q_0=-q_1$. In a representation of the momentum by operators on functions, $\psi(x)$, on Minkowski space, with $x^\pm=x^0\pm x^1$ we have 
$$
q_\mu=-\ri \frac{\partial}{\partial x^\mu}=-i \partial_\mu,\qquad \partial_\pm=\partial_0\pm\partial_1,
$$ and the mass shell is 
$$
\partial_+\partial_-\psi(x)=0, 
$$ 
which gives the  general solution 
$$
\psi(x) =\psi(x^+,x^-)=\psi_+(x^+)+ \psi_-(x^-),
$$
This is the origin of chirality in dimension 2.

 The representation space is the set of square integrable functions on $\R$, $\cL^2( \R)$, with inner product
$$
\langle \tilde f_1,\tilde f_2\rangle=\int_\R \tilde f_1(\phi)^*\tilde{f}_2(\phi)\rd \phi,
$$
 where $\rd\phi $ is the invariant measure on the orbit, corresponding to the Haar measure of $\rSO_0(1,1)$.

Only positive-energy representations $q_0=q>0$ are realized as physical, one-particle states in spacetime. The sign of $q_1$, then, determines the chirality of the representation.  The  negative energy ones are associated, upon quantization, with creation operators of antiparticles, ensuring that the quantum Hamiltonian is bounded from below (see any standard textbook in quantum field theory, for example, \cite{we}). 

It is clear than the same reasoning can be done with fermionic representations and we do not need to develop it here. 
\medskip

{\bf 3) Time-like orbits.} ${{\mathbf q}_t}^2>0$. \label{massorbit} One has two types of orbits, the two disconnected components of the hyperbola defined by 
$$q_0^2-q_1^2= q^2,\qquad q>0.$$ These are given by the sign of $q^0$:
$$q_0=\pm\sqrt{q_1^2+q^2}, \qquad \hbox{so that}\qquad |q_0|\geq q.$$  The little group is again $\Z_2=\{ \pm \id\}$, the base manifold is $\rSO_0(1,1)\cong\R$ and the bundle $E=\R\times \C$.

The orbits with $q_0>0$ can be parametrized as
\be\label{00099}
\begin{pmatrix}q_0\\q_1\end{pmatrix}= q\begin{pmatrix}\cosh\chi\\ \sinh \chi\end{pmatrix},
\ee 
and if $\phi$ is the parameter of a Lorentz transformation $\Lambda_\phi$, the action on the orbit is given simply by 
\be\label{000100}
\Lambda_\phi\begin{pmatrix}q_0\\q_1\end{pmatrix}= q\begin{pmatrix}\cosh(\chi+\phi)\\ \sinh (\chi+\phi)\end{pmatrix}.
\ee 
The orbits with $q_0<0$ can be parametrized as 
\hyphenation{pa-ri-ty}
\be\label{000101}
\begin{pmatrix}q_0\\q_1\end{pmatrix}= q\begin{pmatrix}-\cosh\chi\\-\sinh \chi\end{pmatrix}
\ee
 and the action of the Lorentz group follows similarly, $\chi\to \chi+\phi$. Both orbits are mapped into each other by the operator $TP$ (time inversion  times parity), intertwining the action of the Lorentz group (time inversion alone flips the sign of the parameter). The equivariant functions are formally written as in (\ref{equivariant1,1}), but here the action of the Lorentz transformation $\Lambda_\phi$ is the one described above. One also needs to distinguish between bosonic and fermionic equivariant functions. The action of an element $({\bf a}, v)$ on them is as in  (\ref{transformedsection}). In this case, both translations act, indeed, on the functions.

\hyphenation{asso-cia-ted}

Physically, the positive energy ($q_0>0$) representations can be associated with particles of mass $m=q$. The negative energy ones have a spectrum that is not bounded from below, so they are not admissible as physical particles. Upon quantization, these representations are associated with antiparticles.\medskip

{\bf 4) Space-like orbits.} ${{\mathbf q}_s}^2<0$.\label{spaceorbit} The roles of $q_0$ and $q^1$ are interchanged, so the two disconnected components of the hyperbola are given by the sign of $q^1$:
    $$
    q_0^2-q_1^2=-q^2,\qquad q_1=\pm \sqrt{q_0^2+q^2}.
    $$ 
The parametrization of the orbits $q_1>0$ and $q_1<0$ is similar to \eqref{00099} and  \eqref{000101} for the time-like orbits but interchanging $\cosh\chi$ and $\sinh\chi$. Boosts act also by shifting the  parameter $\chi$. These representations are constructed in the same way as the representations associated to time-like orbits and the formulae do not present formal differences. But now, a Lorentz transformation can change the sign of $q^0$. The energy is not  bounded from below, which signals, physically, an instability. These, otherwise good, UIR of the Poincar\'{e} group cannot be associated with physical particles. For historical reasons, they are called  tachyonic representations, but they are not associated to any `faster than light' particle.

\subsection{Lie algebra representation}
We compute the Lie algebra representation as in Section \ref{lie algebra rep e(2)-sec}. We compare the infinitesimal transformations, (with hermitian generators) in 
\begin{equation} \label{infinitesimalpoincare}
    U(\delta \mathbf{b}, \delta \psi) = e^{\ri \widehat{\mathbf Q} \cdot \delta \mathbf{b}} e^{-\ri K \delta \psi} = \id + \ri \widehat{\mathbf Q} \cdot \delta \mathbf{b} - \ri K \delta \psi + {\mathcal O}(\delta \mathbf{b}, \delta \psi)^2
\end{equation} 
and in \eqref{transformedsection} (notice that this form is valid for all the orbits except ${{\mathbf q}=0}$):
\begin{equation*} \label{boh}
    \tilde{f}^{\tilde g}(\phi) = \tilde{f}(\phi) + i \Lambda_\phi \mathbf{q}_0 \cdot \delta \mathbf{b} - \frac{\partial \tilde{f}}{\partial \phi} \delta \psi + \mathcal{O}(\delta \mathbf{b}, \delta \psi)^2,
\end{equation*}
where the representative of the orbit os $\mathbf{q}_0^t=(q_1,q_2)$, to obtain
\begin{align*}
    \widehat Q_1 &= q_1 \cosh \phi + q_2 \sinh \phi, \\
   \widehat  Q_2 &= q_1 \sinh \phi + q_2 \cosh \phi, \\
    \widehat K &= -\ri \frac{\partial}{\partial \phi},
\end{align*}
From this, it is easy to check the Lie commutators
\begin{equation*}
    [\widehat K, \widehat Q_1] = -i\,\widehat Q_2,\qquad [\widehat K, \widehat Q_2]= -i\,\widehat Q_1, 
    \qquad [\widehat Q_1, \widehat Q_2]=0.
\end{equation*}
By performing the change from hermitian to antihermitian operators, these commutation relations coincide with those given in \eqref{crP(1,1)}.


\subsection{Basis of eigenvectors of $\hat K$}\label{K-eigenfunctions}

It is our purpose now to compute the matrix elements of the infinite dimensional matrix of the transformation (\ref{transformedsection}). 

We consider a basis of eigenvectors of the operator $K=-\ri\partial_\phi$, that is, 
\begin{equation*}
    -\ri \frac{\partial}{\partial \phi} f_p(\phi) = p f_p(\phi) \quad \Rightarrow \quad f_p(\phi) = A_p \re^{\ri p \phi} \quad \text{for } p \in \R,\quad \phi\in \R
\end{equation*}
where  $A_p$ is, in principle, a function of $p$. These functions do not belong to $\cL^2(\R)$, but we can interpret them as  distributions. The appropriate  framework   is the formalism of rigged Hilbert spaces (see Appendix \ref{RHS:appendix}).
 
They are orthogonal:
\begin{equation*}
    \int_\R A_p A^*_{p '} \re^{\ri(p - p ') \phi}\, d\phi = 
   2 \pi \, |A_p|^2 \,\delta(p - p '),
\end{equation*}
so we can choose 
\be\label{001}
f_p (\phi)=\frac 1{\sqrt{2\pi}}\re^{\ri p \phi}, 
\ee
 and we have a continuous orthonormal basis $\{f_p (\phi)\}_{p\in \R}$.

For our purposes, it is enough to consider the group elements in the connected component of the identity. Then, if $\tilde f(\phi)$ is an equivariant section (in any of the orbits except for ${\bf q}_0=0$), the transformation law  under an element of the group $g=({\bf b}, v=\re^{\zeta/2})$ is  (see (\ref{transformedsection})):
$$\tilde f^g(\phi)=\left(U_{{\bf q}_0}({\bf b},\re^{\zeta/2})\tilde f\right)(\phi)=\re^{\ri\Lambda_\phi{\bf q}_0\cdot {\bf b}}\tilde f(\phi-\zeta).$$ Then, the matrix element between $f_p(\phi)$ and $f_{p'}(\phi)$ is
\be \label{generalmatrix}U_{{\bf q}_0}({\bf b},\re^{\zeta/2)})_{p'p}=\langle f_{p'}(\phi)|U_{{\bf q}_0}({\bf b},\re^{\zeta/2})|f_p(\phi)\rangle=
\frac{\re^{-\ri p\zeta}}{2\pi}I(k,\mathbf{q}_0, \mathbf{b}),\ee where $k=p'-p$ and 
 \be I(k,\mathbf{q}_0, \mathbf{b})= \int_\R\rd \phi\,\re^{\ri k\phi}\re^{\ri\Lambda_\phi{\bf q}_0\cdot {\bf b}}.\label{integral}\ee

 We have to evaluate $I(k,\mathbf{q}_0, \mathbf{b})$ in each case. In the first place, we have to distinguish the three kinds of translations, i.e., the time-like ones, $ \mathbf{b}\cdot\mathbf{b}>0$, the space-like ones, $ \mathbf{b}\cdot \mathbf{b}<0$ and the light-like ones, $ \mathbf{b}\cdot \mathbf{b}=0$. We will consider separately $ \mathbf{b}=0$. A generic translation can be written in one of the following forms:
        \[
        \begin{array}{lll}
         \mathbf{b}=\Lambda_{\phi_0}  \mathbf{b}_t ,\qquad  \mathbf{b}_t =(b,0),\;\;b\in \R^\times, \;\phi_0\in \R,  \\[0.3cm]
         \mathbf{b}=\Lambda_{\phi_0}  \mathbf{b}_s, \qquad  \mathbf{b}_s =(0,b),\;\ b\in \R^\times, \;\phi_0\in \R, \\[0.3cm]
           \mathbf{b}=b (1,\delta )^t, \qquad \hskip2.15cm b\in \R^\times, \;\delta=\pm 1.\\[0.3cm]
          \mathbf{b}=0 
           \end{array}\]
    Hence, for the time-like and space-like translations  we have, respectively, 
    \[     
   \Lambda_\phi \mathbf{q} \cdot \mathbf{b}=\Lambda_{\phi-\phi_0} \mathbf{q} \cdot \mathbf{b}_{t/s}.
   \]

   We now choose a representative of each orbit:
   \[
  \mathbf{q}_t=(q,0),  \qquad  \mathbf{q}_s=(0,q),  \qquad  \mathbf{q}_l=q (1,\delta ), \quad q\in \R^\times, \;\delta=\pm 1.
   \]
   Remember that he representations associated to the orbit ${\bf q}=0$ have already been computed in subsection \ref{orbitsso11}. For each orbit we have to consider the three possibilities for ${\bf b}$: 
     
   {\bf 1}. Time-like orbits:
     \[
           \Lambda_\phi \mathbf{q}_t \cdot \mathbf{b}=\left\{  \begin{array}{lll}
       a) \; \Lambda_{\phi-\phi_0}  \mathbf{q}_t \cdot  \mathbf{b}_t =R\,\cosh (\phi-\phi_0),  \\[0.3cm]
        b)\;   \Lambda_{\phi-\phi_0}  \mathbf{q}_t \cdot  \mathbf{b}_s =- R\sinh (\phi-\phi_0),\qquad R=q\,b\in\R^*, \\[0.3cm]
       c) \; \Lambda_{\phi-\phi_0}  \mathbf{q}_t \cdot  \mathbf{b}_l =R\, e^{-\delta\,(\phi-\phi_0)},\;\;\delta=\pm 1.
        \end{array}\right. \]
   
{\bf 2}. Space-like orbits:
     \[
         \Lambda_\phi \mathbf{q}_s \cdot \mathbf{b}=\left\{  \begin{array}{lll}
      a)\;\Lambda_{\phi-\phi_0}  \mathbf{q}_s \cdot  \mathbf{b}_t =R\,\sinh (\phi-\phi_0),  \\[0.3cm]
        b)\; \Lambda_{\phi-\phi_0}  \mathbf{q}_s \cdot  \mathbf{b}_s =R\cosh (\phi-\phi_0),\qquad R=q\,b\in\R^*, \\[0.3cm]
       c)\;\Lambda_{\phi-\phi_0}  \mathbf{q}_s \cdot  \mathbf{b}_l =-\delta\,R\, \re^{-\delta\,(\phi-\phi_0)},\;\;\delta=\pm 1.
        \end{array}\right. \]
   
{\bf 3}. Light-like orbits:
     \[
        \Lambda_\phi \mathbf{q}_l \cdot \mathbf{b}=\left\{  \begin{array}{lll}
      a)\; \Lambda_{\phi-\phi_0}  \mathbf{q}_l \cdot  \mathbf{b}_t =R\,\re^{\delta (\phi-\phi_0)},  \\[0.3cm]
      b)\; \Lambda_{\phi-\phi_0}  \mathbf{q}_l \cdot  \mathbf{b}_s = -\delta\,R\,\re^{\delta (\phi-\phi_0)},\;\; R=q\,b\in\R^*, \\[0.3cm]
       c)\;\Lambda_{\phi-\phi_0}  \mathbf{q}_l \cdot  \mathbf{b}_l =
       \left\{  \begin{array}{lll}
       0 ,\quad {\rm if}\;  \mathbf{q}_l \propto \mathbf{b}_l,\\[0.2cm]
       \delta 2 R\quad {\rm otherwise}.
        \end{array}\right.\end{array}\right. \]
        
For ${\bf b}=0$  this factor becomes trivial. The matrix element is 

$$ U_{{\bf q}_0}(0,\re^{\zeta/2)})_{p'p}=\langle f_{p'}(\phi)|U_{{\bf q}_0}(0,\re^{\zeta/2})|f_p(\phi)\rangle=
\frac{\re^{-\ri p\zeta}}{2\pi}\int_\R\rd \phi\,\re^{\ri k\phi}=\re^{-\ri p\zeta}\delta_D(p-p'),$$ where $\delta_D$ is the Dirac delta, a tempered distribution.

\medskip
   
We now go back to compute the integral (\ref{integral}) for the other cases.

\medskip
Let us consider first  time-like orbits.

\medskip

For the case 1a) we have to solve the integral
$$
    I_{tt}(k,R)= \int_\R \rd\phi \, \re^{-\ri k\phi} \re^{\ri R\cosh (\phi-\phi_0)}.
       $$
By changing the variable $ \chi=\phi-\phi_0$ and decomposing $\re^{-\ri k\chi}=\cos (k\chi) -\sin (k\chi)$ we obtain the integral
$$
    I_{tt}(k,R)=2\re^{-\ri k\phi_0} \int_0^{\infty} \rd\chi \, \cos (k\chi)\,\re^{\ri R\cosh \chi}.
        $$
We now compare with  the  formula 10.32.9 in  Ref.~\cite{NIST} for the modified Bessel functions of second kind $K_\nu(z)$:
$$
  K_\nu(z)= \int_0^{\infty} \rd\chi \, \cosh (\nu\chi)\re^{-z\cosh \chi},\quad  |\arg z |<\pi/2. 
        $$
Taking $\ri k=\nu$ we have $\cos (k\chi)=\cosh(\nu \chi )$. In order to satisfy the constraint on $\arg z$, we have to add a cut off term $\re^{-\varepsilon\, \cosh \chi}$, so $z=\varepsilon-\ri R$, with $\varepsilon>0$.  Hence, we have the integral
 \begin{align*}
    I_{tt}(k,R,\varepsilon) = &\, 2\,\re^{-\ri k\phi_0}\, \int_0^{\infty} \rd\chi \, \cosh (k\chi)\,\re^{-(\varepsilon-\ri R)\,\cosh (\chi)}=
    \\[0.3cm]
   & 2\,\re^{-\ri k\phi_0}\,K_{\ri k}(\varepsilon-\ri R).
   \end{align*}
Taking the limit $\varepsilon \to 0$ we obtain
   $$
    I_{tt}(k,R)= 2\,\re^{-\ri k\phi_0}\,K_{\ri k}(-\ri R).
$$
In formula 10.27.8 of Ref.~\cite{NIST},  the modified Bessel functions of the second kind are related to the Hankel functions:
\be\label{besselhankel}
K_\nu(z)=\frac{\ri\pi} 2\re^{\ri \nu\pi/2}\, H^{(1)}_{\nu}(\ri z),  \quad -\pi\leq \arg (z)\leq \pi/2,
\ee
so we can rewrite our integral in terms of the Hankel functions
 $$
    I_{tt}(k,R)= \ri\pi\re^{-\ri k \phi_0}\,\re^{-k\pi/2}\,H^{(1)}_{\ri k}(R),$$
    and the matrix elements become (\ref{generalmatrix}):
    
   \be \label{matrixtt}U_{tt}({\bf b},\re^{\zeta/2)})_{p'p}=
\frac{i}{2}{\re^{-\ri p\zeta}}\re^{-\ri k \phi_0}\,\re^{-k\pi/2}\,H^{(1)}_{\ri k}(R).\ee

\medskip

The case 1b) corresponds to the integral
$$
    I_{ts}(k,R)= \int_\R \rd\phi \, \re^{-\ri k\phi} \re^{-\ri R\sinh (\phi-\phi_0)},
        $$
that can be transformed into

\begin{align*}
    &I_{ts}(k,R)= \re^{-\ri k\,\phi_0}\,\int_\R \rd\chi \, \re^{-\ri \,k\,\chi} \re^{-\ri \,R\,\sinh \chi}=\\[0.3cm]&2\re^{-\ri k\,\phi_0}\left( \int_{0}^{\infty}\rd\chi \, \cos(k\,\chi) \cos(R\sinh \chi)- \int_{0}^{\infty}\rd\chi \, \sin(k\,\chi) \sin(R\,\sinh \chi)\right).
            \end{align*}

We use now  the following formulas in page 85 of Ref.~\cite{magnus}
\[ \begin{array}{lll}
 \ds K_\nu(x)\,\cos \left(\frac{\pi\,\nu}{2}\right)=\ds \int_0^{\infty} \rd\chi \, \cosh (\nu\chi)\,\cos (x\, \sinh \chi),\\[0.35cm]
\ds K_\nu(x)\,\sin \left(\frac{\pi\,\nu}{2}\right)=\ds \int_0^{\infty} \rd\chi \, \sinh (\nu\chi)\,\sin (x \,\sinh \chi),  \end{array}   
\]
valid for  $ x>0,\;\; -1<\Re (\nu)<1$. In fact, we choose $\nu=\ri k$ and 
we obtain:
$$
I_{ts}(k,R)=  \left\{\begin{array}{lll}
    2 \re^{-\ri k\phi_0}\,\re^{ \pi k/2}\,K_{\ri k}(R),\qquad & R>0,\\[0.3cm]
   2 \re^{-\ri k\phi_0}\,\re^{-\pi\, k/2}\,K_{\ri k}(|R|),\qquad & R<0.
     \end{array}\right.  $$
We can put both together as

$$
I_{ts}(k,R)=
    2\re^{-\ri  k \phi_0}\,\re^{ {\rm sgn}(R)\pi k/2}\,K_{\ri k}(|R|)=  \ri \pi\re^{-\ri k\phi_0}\,\re^{\left({\rm sgn}(R)-1\right)k\pi/2}\,H^{(1)}_{\ri k}(\ri |R|)
  $$
where ${\rm sgn}(R)=R/|R|$. In the last equality we have used the  formula (\ref{besselhankel}). Finally, the matrix element  (\ref{generalmatrix}) becomes:

\be U_{ts}(b, e^{\zeta/2})_{p'p} =
\frac \ri2\,\re^{-\ri p\zeta}\,\re^{-\ri k\phi_0}\,\re^{\left({\rm sgn}(R)-1\right)k\pi/2}\,H^{(1)}_{\ri k}(\ri |R|)\label{matrixts}
\ee

\medskip

The case 1c) corresponds to  light-like translations. The integral to solve is
\be \label{integraltl}
    I_{ tl}(k,R)= \int_\R \rd\phi \, \re^{-\ri k\phi}\, \re^{\ri R\re^{-\delta (\phi-\phi_0)}}
    =\re^{-\ri k\phi_0}\, \int_\R \rd\chi \, \re^{-\ri k\chi}\, \re^{\ri R\,\re^{-\delta \chi}}.
        \ee
        This integral is not absolutely convergent, and we have to regularize it. Let us take first $\delta =-1$, so the integral to solve is
\be I_{tl}(k,R)^{(-)}= \int_\R \rd\chi \, \re^{-\ri k\chi}\, \re^{\ri R\re^{\chi}}.\label{I(kR)-}\ee
        
         For generic values of  $k$ and $R$, we introduce a regulator to control the behaviour $\chi\to +\infty$, the function $\re^{-\varepsilon\re^\chi}$, $\varepsilon>0$. To control the behaviour at $\chi \to -\infty$ we introduce another regulator $\re^{\eta\chi}$, $\eta>0$.
         The result is the integral
        
  \be I_{tl}(k,R)^{(-)\varepsilon}_\eta= \int_\R \rd\chi \, \re^{(\eta-\ri k)\chi}\, \re^{(\ri R-\varepsilon)\re^{\chi}}.\label{convergentintegral}\ee Notice that the second regulator, $\eta$, does affect the behaviour at $+\infty$ because the integrand continues to decrease thanks to the super-exponential decay of  $\re^{-\varepsilon\re^\chi}$. 
  This integral is now absolutely convergent. Performing the change of variable $t=\re^\chi$ we obtain
  
  $$I_{tl}(k,R)^{(-)\varepsilon}_\eta=     
   (\varepsilon -\ri R)^{-(\eta-\ri k)}\Gamma(\eta-\ri k).$$
   The Gamma function has a simple pole at zero, so the limit $\eta\to 0$ behaves badly at one point, $k=0$, which corresponds to the diagonal matrix elements of the representation. 
   
   For $k\neq 0$, and taking the principal branch for the logarithm we obtain:

  \be I_{tl}(k,R)^{(-)}=\begin{cases}R^{\ri k}\re^{k\pi/2}\Gamma(-\ri k),\qquad &R>0,\\[0.3cm]
  2\pi\delta_D(k),\qquad &R=0,\\[0.3cm]
  |R|^{\ri k}\re^{-k\pi/2}\Gamma(-\ri k),\qquad &R<0.\end{cases}\label{I(k,R)-}\ee
  The case $R=0$  ($b=0$) is computed  as $I(k, 0)$ in (\ref{I(kR)-}).
   
   Finally, the matrix element is
   
   \be U_{tl}^{(-)}({\bf b},\re^{\zeta/2})_{p'p}=\frac 1{2\pi}\re^{-\ri p\zeta}\re^{-\ri k\phi_0}I_{tl}(k,R)^{(-)}\label{matrixtl1}\ee
   
   To take into account the point $k=0$ we  can give meaning to this integral as a tempered distribution in $\cS'(\R)$. We consider again the integral (\ref{convergentintegral}) but we now  integrate it directly against the test function $\psi(k)$ in the Schwartz space of functions of rapid decrease (see Appendix \ref{RHS:appendix}):

$$\big( I_{tl}(k,R)_\eta^{(-)\varepsilon},\psi(k)\big)=
\int_{\R}\rd k\int_{\R}\rd \chi\,
\re^{(\eta-\ri k)\chi}\,\re^{(\ri R -\varepsilon)\re^{\chi}}\psi(k).$$ 
The double integral of the modulus becomes
\be\label{integralmodule}\int_\R\rd k\int_\R\rd\chi\,\re^{\eta\chi}\re^{-\varepsilon\re^\chi}|\psi(k)|=\left(\int_\R\rd k|\psi(k)|\right)\left(\int_\R\rd \chi \, \re^{\eta\chi}\re^{-\varepsilon\re^\chi}\right).\ee
The second integral above gives, after using a change of variable $t=\re^\chi$,

$$\int_\R\rd \chi\, \re^{\eta\chi}\re^{-\varepsilon\re^\chi}=\varepsilon^{-\eta}\Gamma(\eta)\,<\, \infty,$$ so (\ref{integralmodule}) is finite and we can use Fubini-Tonelli's Theorem to interchange the order of integration
\begin{align*}
\big( I_{tl}(k, R)_{\varepsilon}^{(-)\eta},\psi(k)\big)=&\int_{\R}\rd\chi\,\re^{\eta\chi}\,\re^{\ri R \e^{\chi}}\re^{-\varepsilon \re^{\chi}}\int_{\R}\rd k\,
\re^{-\ri k\chi}\psi(k)=\\ &2\pi\int_{\R}\rd\chi\,\re^{\eta\chi}\re^{\ri R \re^{\chi}}\,\re^{-\varepsilon \re^{\chi}}\widehat{\psi}(\chi),
\end{align*}
being $\widehat{\psi}(\chi)$ the Fourier transform of $\psi(k)$ 
$$\widehat{\psi}(\chi)=\frac 1{2\pi}\int_{\R}\rd k\,
\re^{-\ri k\chi}\psi(k)$$
(still in the Schwartz space). We have now to take limits when the parameters $\eta$ and $\varepsilon $ go to zero. It is important here the order in which we take the limits. We consider first $\varepsilon$ fixed. By the Lebesgue dominated convergence Theorem \cite{weir}, we can take the limit $\eta\rightarrow 0$ inside the integral. In fact, we have to find a function, independent of $\eta$, such that it is integrable and greater or equal than the modulus of the integrand:
$$M=\re^{\eta\chi}\re^{-\epsilon\re^\chi}|\widehat{\psi}(\chi)|.$$
The modulus of the Schwartz function is integrable. On the other hand, one can observe that
$\re^{-\epsilon\re^\chi}\leq 1$, for fixed $\epsilon$. Let us now consider two cases:

\begin{itemize}
\item
${\chi\leq 0}$. Then 
$\re^{\eta\chi}\leq 1$, and $M\leq |\widehat{\psi}(\chi)|$, which is integrable in the interval $(-\infty, 0]$.

\item  ${\chi>0}$. Since we are taking $\eta\to 0$, we can consider $\eta\leq \eta_0$. Then 
$\re^{\eta\chi}\leq \re^{\eta_0\chi},$ and $M\leq \re^{\eta_0\chi}\re^{-\epsilon\re^\chi} |\widehat{\psi}(\chi)|$ is integrable in $(0,+\infty)$ because of the superexponential factor. 

\end{itemize}
So the function defined piecewise in the two intervals is integrable and does not depend on $\eta$. This implies that we can take the limit inside the integral and 
$$\lim_{\eta\rightarrow 0^+}( I_{tl}(k, R)_{\varepsilon}^{(-)\eta},\psi(k))=2\pi\int_{\R}\rd\chi\,\re^{\ri R \re^{\chi}}\re^{-\varepsilon \re^{\chi}}\widehat{\psi}(\chi)$$

For the second limit, we notice that
$$\re^{-\varepsilon \re^{\chi}}|\widehat{\psi}(\chi)|\leq |\widehat{\psi}(\chi)|,$$ which is integrable, so we can use again the dominated convergence Theorem and we get that
$$\lim_{\varepsilon\to 0^+}\lim_{\eta\rightarrow 0^+}\langle I_{tl}(k, R)^{(-)\varepsilon}_{ \eta},\psi(k)\rangle=2\pi\int_{\R}\rd\chi\,\re^{\ri R \re^{\chi}}\widehat{\psi}(\chi).$$
This integral is absolutely convergent.  Defining the distribution
$$
\big(D^{(-)}_R(k),\psi(k)\big):=2\pi\int_{\R}\rd\chi\,\re^{\ri R \re^{\chi}}\widehat{\psi}(\chi),
$$ 
we can finally write  the matrix element as
\be U_{tl}^{(-)} ({\bf b},\re^{\zeta/2})_{p'p}=\re^{-\ri p\zeta}\re^{-\ri k\phi_0}D^{(-)}_R(k).\label{matrixtl2}\ee

\medskip
It remains to consider, for the time-like orbits and light-like translations, the case with $\delta=+1$ (see(\ref{integraltl})). The integral to consider is
$$ I_{tl}(k,R)^{(+)}= \int_\R \rd\chi \, \re^{-\ri k\chi}\, \re^{\ri R\re^{-\chi}}.$$ Now the rapid oscillations are in $\chi\to -\infty$. We can  regularize it in a similar way
\be I_{tl}(k,R)_{\varepsilon}^{(+)\eta}= \int_\R \rd\chi \, \re^{(-\eta-\ri k)\chi}\, \re^{(\ri R-\varepsilon)\re^{-\chi}},\label{I(k,R)+}\ee for $\eta,\varepsilon >0$. The term $\re^{-\eta\chi}$ controls the behaviour when $\chi\to \infty$ and $\re^{-\varepsilon\re^{-\chi}}$ controls the behaviour when $\chi\to -\infty$. Performing the change of variable $t=\re^{-\chi}$ we arrive to the expression

$$I_{tl}(k,R)^{(+)\eta}_{\varepsilon}
=
(\varepsilon - iR)^{-(ik+\eta)}\,\Gamma(ik+\eta).
$$
For $k\neq 0$, taking the limits and choosing the principal branch for the logarithm we get:
$$I_{tl}(k,R)^{(+)}=(-\ri R)^{-\ri k}\Gamma(\ri k)=\begin{cases}R^{-\ri k}\re^{-k\pi/2} \qquad &R>0,\\[0.3cm]
2\pi\delta(k) \qquad &R=0,\\[0.3cm]
|R|^{-\ri k}\re^{k\pi/2}\qquad &R<0.\end{cases}$$ The matrix element then becomes
\be U^{(+)}_{tl}({\bf b},\re^{\ri\zeta})_{p'p}=\frac 1{2\pi}\re^{-\ri p\zeta}\re^{-\ri k\phi_0}I_{tl}(k,R)^{(+)}.\label{matrixtl3}\ee
To take into account  $k=0$ we integrate (\ref{I(k,R)+}) against a test function and an argument similar to the case $\delta=-1$ can be made. The result is 
$$
\lim_{\varepsilon\to 0^+}\lim_{\eta\to 0^+}
\big( I_{tl}(k,R)^{(+)\eta}_\varepsilon,\psi(k)\big)
=
2\pi \int_\R \rd\chi\,
e^{iRe^{-\chi}}
\widehat{\psi}(\chi).
$$
This integral is absolutely convergent. We define the distribution
$$
\big( D_R^{(+)}(k),\psi(k)\big)
:=
2\pi \int_\R \rd\chi\,
\re^{iR\re^{-\chi}}
\widehat{\psi}(\chi).
$$
Therefore,
$$
I_{tl}^{(+)}(k,R)=D_R^{(+)}(k)
$$
as a tempered distribution, and the matrix element is

\be U^{(+)}_{tl}({\bf b},\re^{\ri\zeta})_{p'p}=\re^{-\ri p\zeta}\re^{-\ri k\phi_0}D_R^{(+)}(k).\label{matrixtl4}\ee
The expressions (\ref{matrixtl1}), (\ref{matrixtl2}), (\ref{matrixtl3}) and (\ref{matrixtl4}) complete the calculation of the matrix elements for the time-like orbits and light-like translations. 

\bigskip

We consider now space-like orbits.

\medskip

The case 2a) of the space-like orbits is  very similar to the case 1b) of the time-like orbits. The result is:
\be U_{st}({\bf b},\re^{\ri\zeta})_{p'p}=\frac \ri2\,\re^{-\ri p\zeta}\,\re^{-\ri k\phi_0}\,\re^{-\left({\rm sgn}(R)+1\right)k\pi/2}\,H^{(1)}_{\ri k}(\ri |R|)\label{matrixst}\ee

\medskip

The case 2b) is identical to 1a). The result is:

\be \label{matrixss}U_{ss}({\bf b},\re^{\zeta/2)})_{p'p}=
\frac{i}{2}{\re^{-\ri p\zeta}}\re^{-\ri k \phi_0}\,\re^{-k\pi/2}\,H^{(1)}_{\ri k}(R).\ee

\medskip

As for the time-like orbits we have in 2c) two different cases, according to the sign of $\delta$. The integrals turn out to be the same than in the time-like case, up to a sign. More precisely, 
\be U^{(\pm)}_{sl}({\bf b},\re^{\zeta/2)})_{p'p}=\mp U^{(\pm)}_{tl}({\bf b},\re^{\zeta/2)})_{p'p},\label{matrixsl}\ee 
which are given by formulae (\ref{matrixtl1}), (\ref{matrixtl2}), (\ref{matrixtl3}) and (\ref{matrixtl4}).

\bigskip

We consider now light-like orbits. The integrals that appear are all computed already or are trivial. We just write the results.
\hyphenation{na-tu-ra-lly}

\medskip

Case 3a)

 \be
 U^{(\pm)}_{lt}({\bf b},\re^{\zeta/2)})_{p'p}=\pm\,U^{(\mp)}_{tl}({\bf b},\re^{\zeta/2)})_{p'p}. \label{matrixlt}
 \ee
\medskip

Case 3b)

\be
U^{(\pm)}_{ls}({\bf b},\re^{\zeta/2)})_{p'p}=\mp U^{(\mp)}_{tl}({\bf b},\re^{\zeta/2)})_{p'p}.
\label{matrixls}
\ee

\medskip

Case 3c) If the translation ${\bf q_l}\propto {\bf b}_l$ then $$U_{ll}^\propto({\bf b},\re^{\zeta/2)})_{p'p}=0. $$ In the other cases the integral becomes, essentially, a Dirac delta:
\be
U^{(\delta)}_{ll}({\bf b},\re^{\zeta/2)})_{p'p}=\re^{-\ri p \zeta}\,\re^{\delta \ri 2 R}\delta_D(k). \label{matrixll}
\ee

\section*{Conclusions}
\hyphenation{re-la-ti-vis-tic}

In this work we have constructed explicitly the UIR of the 2D Euclidean and Poincar\'e groups, by means of Mackey's theory of induced representations. The simplicity of the corresponding little groups allows a complete  implementation of the induced
representation program, including the explicit description of momentum
orbits, equivariant wavefunctions, invariant measures, infinitesimal
generators, and matrix elements of the representation operators.

For the Euclidean group and its double cover, the representations are naturally classified according to trivial and non-trivial momentum orbits, with a clear distinction between bosonic and fermionic sectors, arising from
the Spin covering. The corresponding infinite-dimensional representations
were realized explicitly in terms of equivariant functions on the orbit
space, and their matrix elements were computed in closed form in terms of
Bessel functions.

 For the Poincar\'e group in dimension $1+1$, the Lorentzian structure of
momentum space leads to a richer orbit decomposition, including time-like,
space-like, and light-like orbits. The associated induced representations
were constructed explicitly for all cases. The corresponding matrix elements
involve modified Bessel and Hankel functions and, in some situations,
 appear naturally as tempered distributions. This required the introduction
of Rigged Hilbert Spaces as the appropriate functional framework for the
treatment of generalized eigenvectors and continuous bases.


Beyond the explicit calculations themselves, the present work highlights the
deep interplay among harmonic analysis on Lie groups, induced
representations, Clifford algebras, special functions, and geometric methods
in mathematical physics. In particular, the appearance of classical special
functions as matrix elements of group representations illustrates the close
relation between representation theory and analysis on homogeneous spaces.

The two-dimensional setting considered here provides a particularly useful
laboratory where Mackey's construction can be implemented in complete
detail. In this sense, the present analysis may serve both as a concrete
reference example for induced representation theory and as a starting point
for more general constructions involving non-compact Lie groups and their
applications in relativistic quantum theory.

\section*{Acknowledgements}
M. A. Lled\'{o} wants to thank the Departamento de F\'{\i}sica Te\'{o}rica, At\'{o}mica y \'{O}ptica  of Universidad de Valladolid, for its kind hospitality during the realization of this work. \\ 
\noindent This work is supported by the Spanish Grants PID2020-116567GB-C21, PID2023-001292-S and CEX2023-001292-S,
funded by\\
\noindent MCIU/AEI/10.13039/501100011033.\\
 \noindent This work was supported by Horizon Europe EU projects: \\
 \noindent MSCA-SE CaLIGOLA, Project ID: 101086123,\\
\noindent MSCA-DN CaLiForNIA, Project ID: 101119552.\\
\noindent This article is based upon work from COST Action CaLISTA CA21109 supported by COST (European Cooperation in Science and Technology), www.cost.eu.
\hyphenation{hos-pi-ta-li-ty}
\hyphenation{rea-li-za-tion}

\section*{Appendices}

\appendix 
\section{ Clifford algebras and Spin groups} \label{app:clifford} In this Appendix, we perform a quick review of some concepts on Clifford algebras. More detailed expositions of these classical results can be found in many places (see, for example, Refs. \cite{lm,vasusy}). This brief report is mainly based in Refs. \cite{vasusy,ga}.
 We  only state the results that we will actually use.

\subsection{Complex Clifford algebras}
Let $V$ be a finite dimensional, complex vector space of dimension $D$. Let
$$\begin{CD}V\times V@>\varphi>>\C\end{CD}$$
be a non degenerate, symmetric, bilinear form and let $Q(v)=\varphi(v,v)$ be the associated quadratic form on $V$.

Associated to such $\varphi$, a Clifford algebra, $\cCl(V)$, can be defined as the quotient of the tensor algebra of $V$, $T(V)$, by the ideal
$$\cI= \left(u\otimes v+v\otimes u-2\varphi(u,v)\cdot 1\right),\qquad u,v\in V.$$ One can always find an orthonormal basis $\{e_\mu\}_{\mu=1}^D$ of $V$ where $\varphi(e_\mu,e_\nu)=\delta_{\mu\nu}$, so the Clifford algebras for different $V$'s, with the same dimension,  are all isomorphic. We will denote them often by $\cCl(D)$. The dimension of the Clifford algebra is $2^D$.

While the tensor algebra is $\Z$-graded, the Clifford algebra conserves only  a $\Z_2$-grading, i.e., 
$\cCl(V)=\cCl(V)^+ \oplus \cCl(V)^-$, where the even part, $\cCl(V)^+$, is  spanned by tensors of even rank and the odd part, $\cCl(V)^-$, is spanned by tensors of odd rank.  It is, in fact, a superalgebra. 
The vector space $V$ sits inside the Clifford algebra.

 In even dimension,  $D=2m$, $\cCl(D)\cong\rEnd(S)$, where $S$ is a vector space of dimension $2^{m}$ (as a superalgebra, the dimension is $2^{m-1|m-1}$).

 The odd case, $D=2m+1$, is a bit more complicated. Let  $F=\C[\epsilon]$ be a superalgebra with $\epsilon$ odd and $\epsilon^2=1$. It is a super division algebra\footnote{A super division algebra is a superalgebra whose non zero homogeneous elements are invertible.} and it is isomorphic to the center of $\cCl(D)$, denoted as $Z$. Then, for some vector space $S_0$, of dimension  $2^m$, $\cCl(D)=\rEnd(S_0)\otimes Z$. We have also that $\cCl(D)^+\simeq \rEnd(S_0)$.

 For arbitrary dimension, we consider the group of invertible elements in $\cCl(V)$, $\cCl(V)^\times$. We define the {\it  Clifford group} as
 $$\Gamma=\{u\in \cCl(V)^\times, u\hbox{ homogeneous } |\; uVu^{-1}\subset V\}$$ and the {\it even Clifford group} as
$$\Gamma^+=\{u\in {\cCl(V)^\times}^+ |\; uVu^{-1}\subset V\}.$$ For each $u\in \Gamma$ there is an action
\be\begin{CD}V@>\alpha(u)>>V\\
v@>>>(-1)^{|u|}uvu^{-1},\end{CD}\label{alpha}
\ee being $|u|$ the parity of $u$.
Since
$$Q(uvu^{-1})\cdot 1=(uvu^{-1})^2=Q(v)\cdot 1,\qquad v\in V,$$
The map $\alpha$ is a surjective homomorphism { of groups} from $\Gamma$ to $\rO(V)$. Notice that $\alpha(u)=\alpha(-u)$.   We are going to determine the Spin group as a subgroup of $\Gamma^+$.

The {\it canonical} or {\it principal antiautomorphism}, $\beta$, is the unique antiautomorphism of the ungraded Clifford algebra which is the identity on $V$:
\[
\beta(x_1\cdots x_r)=(x_r\cdots x_1),\qquad x_{i}\in V.
\]
Then the map $x\mapsto x\beta(x)$ is an homomorphism of $\Gamma$ into $\C^\times \cdot 1$. Consider its restriction to $\Gamma^+$. Then, the kernel of this map is an analytic subgroup of $\cCl(V)^\times$ that is a double cover of $\rSO(V)$. This is the Spin$(V)$ group:
\be\label{000505}
\rSpin(V)=\{x\in{\Gamma^+}\;|\;x\beta(x)=1\}.
\ee 
We have then the exact sequence
$$\begin{CD}1@>>>\{\pm 1 \}@>>> \rSpin(V)@>\alpha>> \rSO(V)@>>>1.\end{CD}$$

\subsection{Real Clifford algebras}

Over the real numbers, there are different Clifford algebras with the same dimension. These depend on the signature of the bilinear form. In an orthonormal basis, the bilinear form is represented by a diagonal matrix
$$\eta=\diag(+1,\cdots (r \hbox{ times})\cdots,+1, -1, \cdots (s \hbox{ times})\cdots,-1).$$ The dimension $D=s+t$ and the {\it signature} $\rho=s-t$ have the same parity, i.e., $(-1)^{D}=(-1)^{\rho}$. We will denote the real Clifford algebras with two arguments, $\cCl(s,t)$.

 The Clifford algebras are isomorphic to matrix algebras, either real, complex or  quaternionic, depending on the signature (this and other properties of the Clifford algebras are listed, for example,  in \cite{dflv}). For $D=2$ we have three possibilities:
\begin{enumerate}
\item $\rho=2$, then $\cCl(2,0)\cong \R(2)$, i.e., the real matrices of dimension 2,
\item $\rho =0$, then $\cCl(1,1)\cong  \R(2)$,
 \item $\rho=-2$, then $\cCl(0,2)\cong \bH$.   The quaternions $\qq\in \bH$ can be represented by  2$\times$2 complex matrices $M_{\qq}$ satisfying
     $$
     M_{\qq}^*=-\Omega M_{\qq}\Omega, \qquad \Omega=\begin{pmatrix}0&1\\-1&0\end{pmatrix}.
     $$
     The algebra of these $M_{\qq}$ matrices (the quaternions) is denoted sometimes by $ \bH(2)$.
\end{enumerate}

In particular, notice that $\cCl(2,0)\ncong\cCl(0,2)$.

  The image of the $e_\mu$ under the isomorphisms mentioned above is a set of {\it gamma matrices}, $\{\gamma_\mu\}_{\mu=1}^D$. They satisfy
 the relations
$$
 \{\gamma_\mu, \gamma_\nu\}=\gamma_\mu\gamma_\nu+\gamma_\nu\gamma_\mu=2\eta_{\mu\nu}\,\id.
$$ 
The gamma matrices have (real or complex, depending on  $\rho$) dimension $2^m$, where $m=D/2$ for $D$ even and $m=(D-1)/2$ for $D$ odd.
The Spin group, $\rSpin(r,s)$, is a real form of the complex one, given by the fixed points of a certain conjugation. It maps surjectively onto $\rSO_0(s,t)$, the connected component of the identity of the orthogonal group

One can prove that the Clifford algebra elements
 \be M_{\mu\nu}=\frac 12(e_\mu e_\nu-e_\nu e_\mu)\label{Lorentzspin}
 \ee
 satisfy the commutation rules of the  Lie algebra 
 $\fso(s,t)$, 
  \be
  [M_{\mu\nu}, M_{\rho\sigma}]=\eta_{\mu\rho}M_{\nu\sigma}-\eta_{\mu\sigma}M_{\nu\rho}-\eta_{\nu\rho}M_{\mu\sigma} +
\eta_{\nu\sigma}M_{\mu\rho},\label{CRLorentz}
\ee
Moreover, one can prove that this is the Lie algebra of the Spin$(s,t)$ group. The same works over the complex field.
Notice that  $\fso(s,t)\simeq \fso(t,s)$, so we can choose which Clifford algebra to use of the two available for the same orthogonal group.

For $D$ even,
 the gamma matrices representation splits under $\fso(s,t)$ into two invariant subspaces, usually denoted as $S^\pm$, of dimension $2^{D/2-1}$.  For $D$ odd, it remains irreducible. These are the spinor representations, the irreducible representations of $\fso(s,t)$ whose highest weights are the fundamental weights corresponding to the right extreme nodes of the Dynkin diagram. In the physics language these are {\it Weyl spinors}.  They are representations of the algebra that do not lift to representations of the orthogonal group, but instead, of its  Spin group.

 \section{ Spin groups in dimension 2}\label{spingroups:appendix}
\subsection{Spin group in Euclidean signature, $\rSpin(2)$}\label{spingroupE(2)}
The real Clifford algebra in dimension $D=2$ ($V=\R^2$) and signature $\rho=2$, is
$$
\{e_0,e_0\}=2\cdot 1,\qquad \{e_1,e_1\}=2\cdot 1,\qquad \{e_0,e_1\}=0.$$
A basis for it is
$$\cCl(2,0)=\{1,e_0,e_1,e_0e_1\}.$$ Since it is isomorphic to $\R(2)$, one can find a set of real gamma matrices, for example,
\be
\id=\begin{pmatrix}1&0\\0&1\end{pmatrix},\;
 \gamma_0=\begin{pmatrix}0&1\\1&0\end{pmatrix},\;
\gamma_1=\begin{pmatrix}1&0\\0&-1\end{pmatrix},\gamma_0\gamma_1=
\begin{pmatrix}0&-1\\1&0\end{pmatrix}.\label{gamma2,0}
\ee 
Then, the conjugation defining this real form is the standard complex conjugation.

The $\Z_2$-grading of  $\cCl(2 ,0)$ is
$$
\cCl^+(2 ,0)=\rspan\{\id,e_0e_1\},\qquad \cCl^-(2 ,0)=\rspan\{e_0,e_1\}\simeq V.
$$

 Given an arbitrary element in $x\in \cCl(2,0)$, $x=a\,\id+be_0+ce_1+de_0e_1$, it is invertible if and only if
$$
a^2-b^2-c^2+d^2\neq 0.
$$ 
For the homogeneous elements, the inverses read:
$$
(a \cdot 1+d e_0e_1)^{-1}=\frac{1}{a^2+d^2}(a 1-de_0e_1),\quad (be_0+ce_1)^{-1}=\frac 1{b^2+c^2}(be_0+ce_1).
$$
In this simple case, all the invertible, homogeneous elements belong to the Clifford group, $\Gamma(2,0)$. 

The map $\alpha$ (\ref{alpha}) sends each element of the Clifford group to a rotation in $\rO(2)$.

  The Spin group, $\rSpin(2):=\rSpin(2,0)$, is given by 
  \[
  \rSpin(2)=\left\{ x=a\,\id+de_0e_1 \,|\, x\beta(x)=a^2+d^2=1, \;\;a,d\in \R\right\}
  \]
  In the representation of gamma matrices (\ref{gamma2,0}) they are of the form
 $$
 x=\begin{pmatrix}a&-d\\d&a\end{pmatrix},\qquad a^2+d^2=1.
 $$
%


 Note that for $x=a\cdot 1+de_0e_1\neq 0 \in  \cCl(2,0)$, we have that $\alpha(x)$ \eqref{alpha} has the explicit expression
$$
\alpha(x)=\frac 1{a^2+d^2}\,\begin{pmatrix}a^2-d^2&2ad\\-2ad&a^2-d^2\end{pmatrix}=\begin{pmatrix}a^2-d^2&2ad\\-2ad&a^2-d^2\end{pmatrix}.
$$
Hence, there is a group homomorphism
\be
\begin{CD}\rSpin(2)@>\varphi>>\rSO(2)\\[0.2cm]
\begin{pmatrix}a&-d\\d&a\end{pmatrix}@>>>\begin{pmatrix}a^2-d^2&-2ad\\2ad&a^2-d^2\end{pmatrix}
\end{CD}\label{homE(2)Ap}
\ee 
that, choosing  a convenient parametrization is simply:
$$
\begin{CD}\begin{pmatrix}\cos \theta/2&-\sin\theta/2\\\sin\theta/2&
\cos \theta/2\end{pmatrix}@>\varphi>>\begin{pmatrix}\cos \theta
&-\sin\theta\\\sin\theta&\cos\theta\end{pmatrix}.\end{CD}
$$ 
It is then clear that the Spin group is a double cover of the special orthogonal group.

 On the other hand for $x=be_0+ce_1\in  \cCl(2,0)$ with $b^2+c^2=1$, we obtain through $\alpha$  \eqref{alpha} also a rotation:
$$
\alpha(x)(v)=\frac 1{b^2+c^2}\begin{pmatrix}b^2-c^2&2bc\\2bc&-b^2+c^2\end{pmatrix}
\begin{pmatrix}v^0\\v^1\end{pmatrix},
$$ 
with 
 $\det\alpha(x)=-1$, i.e., the other connected component of  $\rO(2)$.


\subsection{Spin group in Minkowskian signature, $\rSpin(1,1)$}\label{spingroupP(1,1)}

The real Clifford algebra in dimension $D=2$ ($V=\R^2$)  and signature $\rho=0$, $\cCl(1,1)$, is
$$
\{e_0,e_0\}=2\cdot 1,\qquad \{e_1,e_1\}=-2\cdot 1,\qquad \{e_0,e_1\}=0.$$
A basis for  $\cCl(1,1)$ is
$=\{1,e_0,e_1,e_0e_1\}$. Since it is also isomorphic to $\R(2)$, one can find a set of real gamma matrices, for example,
\be\label{gamma1,1}
\id=\begin{pmatrix}1&0\\0&1\end{pmatrix},\; \gamma_0=\begin{pmatrix}0&1\\1&0\end{pmatrix},\; 
\gamma_1=\begin{pmatrix}0&1\\-1&0\end{pmatrix},\; 
\gamma_0\gamma_1= \begin{pmatrix}-1&0\\0&1\end{pmatrix}.
\ee 
Then, the conjugation defining this real form is the standard complex conjugation.

The $\Z_2$-grading of $\cCl(1,1)$ is
$$
\cCl^+(1,1)=\rspan\{\id,e_0e_1\},\qquad \cCl^-(1,1)=\rspan\{e_0,e_1\}\simeq V.
$$

An arbitrary element  $x\in \cCl(1,1)$, $x=a\,\id+be_0+ce_1+de_0e_1$, is invertible if and only if
$$
(a^2-b^2)^2-(c^2-d^2)^2\neq0.
$$ 
The inverses of the homogeneous elements are
$$
(a\cdot 1+de_0e_1)^{-1}=\frac{1}{a^2-d^ 2}(a\cdot 1-de_0e_1),\qquad (b\e_0+ce_1)^{-1}=\frac{1}{b^2-c^ 2}(be_0+ce_1).
$$
Also in this  case, all the invertible, homogeneous elements belong to the Clifford group, $\Gamma(1,1)$. They are sent by $\alpha$ (\ref{alpha}) to the orthogonal  group $\rO(1,1)$.

  The Spin group, $\rSpin(1,1)$, is 
  \begin{align*}
  \rSpin(1,1)=&\left\{
  x=a\cdot 1+de_0e_1\;|\; x\beta(x)=a^2-d^2=(a-d)(a+d)=1;\right.\\ &\left. a,d\in\R\right\}.
  \end{align*}
  In the representation of the gamma matrices (\ref{gamma2,0}) its elements are of the form
 $$x=\begin{pmatrix}a-d&0\\0&a+d\end{pmatrix},\qquad a^2-d^2=1.$$

 For $x=a\cdot 1+de_0e_1\in  \cCl(1,1)$ such that $a^2-d^2\neq 0$, we have
$$\alpha(x)=\frac{1}{a^2-d^2}\begin{pmatrix}a^2+d^2&-2ad\\-2ad&a^2+d^2\end{pmatrix}.$$
 The transformation $\alpha(x)$ has determinant 1, so it belongs to $\rSO(1,1)$. Moreover, if $a^2-d^2=1$, it belongs to $\rSO_0(1,1)$.
 
 A convenient parametrization   of $\rSpin(1,1)$ is 
 $$\begin{pmatrix}\pm\re^{\chi/2}&0\\0&\pm\re^{-\chi/2}\end{pmatrix},\qquad \chi\in \R,
 $$
 with
 $$a=\pm\cosh\frac\chi 2,\qquad d=\mp\sinh\frac\chi 2.$$ 
 So $\rSpin(1,1)=\R^\times\cong\Z_2\times \R^\times_+$ \footnote{This case and $\rSpin(1)=\{\pm 1\}=\Z_2$ are the only cases where the spin group is disconnected.}. The homomorphism  $\alpha$ sends $x$ to 
 $$\alpha(x)=\begin{pmatrix}\cosh \chi&\sinh \chi\\\sinh \chi&\cosh \chi\end{pmatrix}.$$
 The orthogonal group $\rSO_0(1,1)$ is isomorphic to $\R^+$ via $\alpha(x)\rightarrow \re^{\chi}$, so the double covering property 
 $$\pm\re^{\chi/2}\longrightarrow \re^{\chi}$$ is manifest. An account of this computation of the Spin group is given, for example, in \cite{vasusy}.

 For $x=be_0+ce_1\in \cCl(1,1)$ such that  $b^2-c^2\neq 0$ we have
 $$\alpha(x)=-\frac{1}{b^2-c^2}\begin{pmatrix}b^2+c^2&-2bc\\2bc&-b^2-c^2\end{pmatrix}.$$
 This is a transformation of $\rO(1,1)$ with determinant~$-1$.

\section{Rigged Hilbert Spaces}\label{RHS:appendix}

It is well known that the Hilbert space formalism, while central to the standard formulation of Quantum Mechanics, is not sufficient to accommodate all physically relevant structures, even in the non-relativistic setting. In particular, the Dirac formalism \cite{dirac} relies on operators with continuous spectra whose generalized eigenvectors cannot be represented as elements of the Hilbert space of square-integrable wave functions.  For other examples where there appear this kind of difficulty, see for instance \cite{bohm-gadella,bohm-gadella-1}.
\hyphenation{re-pre-sen-ted}

A mathematically natural framework that incorporates the Hilbert space formalism while resolving these limitations is provided by rigged Hilbert Spaces (RHS). This construction was introduced by Gelfand and his collaborators \cite{gelfand}, in the context of the spectral theory of \hyphenation{co-lla-bo-ra-tors}
self-adjoint operators, and further developed through the nuclear spectral theorem, established by Gelfand and Maurin \cite{maurin}. Its relevance to Quantum Mechanics was emphasized in Roberts \cite{roberts}, Antoine \cite{antoine}, Melsheimer \cite{melheimer}, Bohm
\cite{bohm} and de la Madrid \cite{delamadrid}. Recently, in a series of papers Celeghini, Gadella and del Olmo have studied the close connection among Lie groups,  special functions  and RHS (see for instance \cite{olmo2016,olmo2019,olmo2024}). 

A rigged Hilbert Space, or Gelfand triplet, is defined as a triple of spaces
\[
\Phi \subset \mathcal H \subset \Phi^{\times},
\]
where $\mathcal H$ is an infinite-dimensional separable Hilbert space, $\Phi$ is a dense subspace of $\cH$ endowed with a locally convex topology that is strictly finer than the topology induced by $\cH$, and $\Phi^{\times}$ denotes the space of continuous antilinear functionals on $\Phi$.

The refinement of the topology on $\Phi$ ensures that, every convergent sequence in $\Phi$, converges in 
$\cH$, whereas the converse implication does not, generally, hold. Consequently, the dual space $\Phi^{\times}$ is strictly larger than $\cH$, which is self-dual by virtue of the Riesz representation theorem. Elements of $\Phi^{\times}$ act as continuous mappings from $\Phi$ into $\mathbb{C}$.

A central property of the RHS framework is the natural extension of operators. Let $A$ be a densely defined linear operator on $\cH$, such that $\Phi \subset \mathcal{D}(A)$ (the domain of $A$) and $A\Phi \subset \Phi$. In this case, $\Phi$ is said to be invariant under $A$, and $A$ admits a unique extension $A^{\times}$ to $\Phi^{\times}$ defined through the duality relation
\[
\langle A^{\times}F \mid \varphi \rangle := \langle F \mid A\varphi \rangle,
\qquad \forall\, \varphi \in \Phi,\ \forall\, F \in \Phi^{\times}.
\]
If the operator $A$ is continuous with respect to the topology of $\Phi$, then its extension $A^{\times}$ is continuous on $\Phi^{\times}$, when the latter is equipped with the weak$^\ast$ topology.

The locally convex topology of $\Phi$ is typically generated by a countable family of seminorms $\{\|\cdot\|_n\}_{n\in\mathbb{N}}$. A linear operator $A:\Phi \to \Phi$ is continuous if and only if, for each $n$, there exist a constant $C_n > 0$ and a finite subset $\{p_1,\dots,p_k\}$ of indices such that \cite{simon}
\[
\|A\varphi\|_n \leq C_n \sum_{j=1}^{k} \|\varphi\|_{p_j},
\qquad \forall\, \varphi \in \Phi.
\]
An analogous condition characterizes the continuity of linear or antilinear functionals on $\Phi$.


\end{document}

%% file: bundles.tex
\begin{picture}(0,0)%
\includegraphics{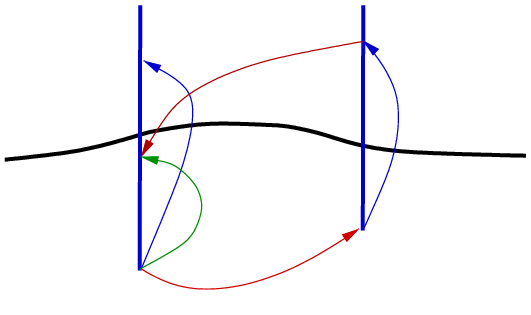}%
\end{picture}%
%
%
\setlength{\unitlength}{3947sp}%
\begingroup\makeatletter\ifx\SetFigFont\undefined%
\gdef\SetFigFont#1#2#3#4#5{%
  \reset@font\fontsize{#1}{#2pt}%
  \fontfamily{#3}\fontseries{#4}\fontshape{#5}%
  \selectfont}%
\fi\endgroup%
\begin{picture}(4240,2510)(975,-1722)
\put(990,-384){\makebox(0,0)[lb]{\smash{{\SetFigFont{12}{14.4}{\rmdefault}{\mddefault}{\updefault}{\color[rgb]{0,0,0}$G/H$}%
}}}}
\put(1748,-1505){\makebox(0,0)[lb]{\smash{{\SetFigFont{12}{14.4}{\rmdefault}{\mddefault}{\updefault}{\color[rgb]{0,0,.82}$[g']$}%
}}}}
\put(1504,242){\makebox(0,0)[lb]{\smash{{\SetFigFont{12}{14.4}{\rmdefault}{\mddefault}{\updefault}{\color[rgb]{0,0,.82}$\varphi([g'])$}%
}}}}
\put(3941,-1076){\makebox(0,0)[lb]{\smash{{\SetFigFont{12}{14.4}{\rmdefault}{\mddefault}{\updefault}{\color[rgb]{0,0,.82}$[g^{-1}g']$}%
}}}}
\put(3934,479){\makebox(0,0)[lb]{\smash{{\SetFigFont{12}{14.4}{\rmdefault}{\mddefault}{\updefault}{\color[rgb]{0,0,.82}$\varphi([g^{-1}g'])$}%
}}}}
\put(1093,-670){\makebox(0,0)[lb]{\smash{{\SetFigFont{12}{14.4}{\rmdefault}{\mddefault}{\updefault}{\color[rgb]{.69,0,0}$g\varphi([g^{-1}g'])$}%
}}}}
\put(2584,-916){\makebox(0,0)[lb]{\smash{{\SetFigFont{12}{14.4}{\rmdefault}{\mddefault}{\updefault}{\color[rgb]{0,.56,0}$\varphi'$}%
}}}}
\put(2473,-421){\makebox(0,0)[lb]{\smash{{\SetFigFont{12}{14.4}{\rmdefault}{\mddefault}{\updefault}{\color[rgb]{0,0,.82}$\varphi$}%
}}}}
\put(4168,-163){\makebox(0,0)[lb]{\smash{{\SetFigFont{12}{14.4}{\rmdefault}{\mddefault}{\updefault}{\color[rgb]{0,0,.82}$\varphi$}%
}}}}
\put(2893,386){\makebox(0,0)[lb]{\smash{{\SetFigFont{12}{14.4}{\rmdefault}{\mddefault}{\updefault}{\color[rgb]{.82,0,0}$g$}%
}}}}
\put(2983,-1648){\makebox(0,0)[lb]{\smash{{\SetFigFont{12}{14.4}{\rmdefault}{\mddefault}{\updefault}{\color[rgb]{.82,0,0}$g^{-1}$}%
}}}}
\end{picture}%